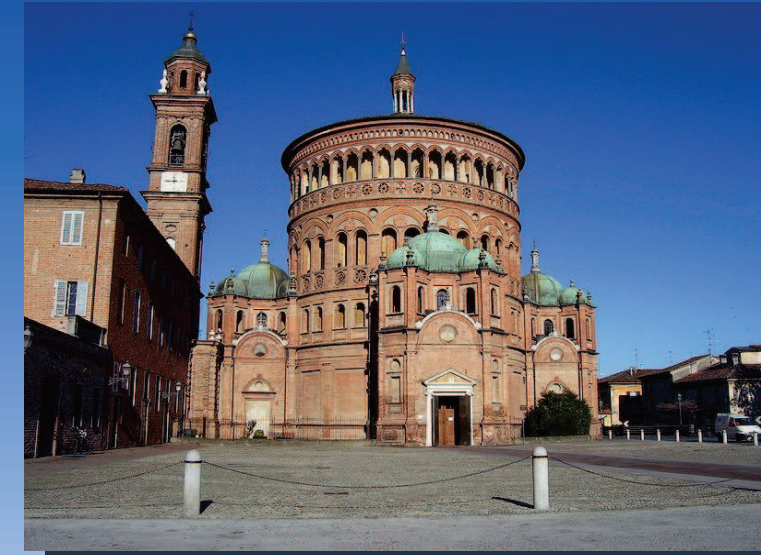
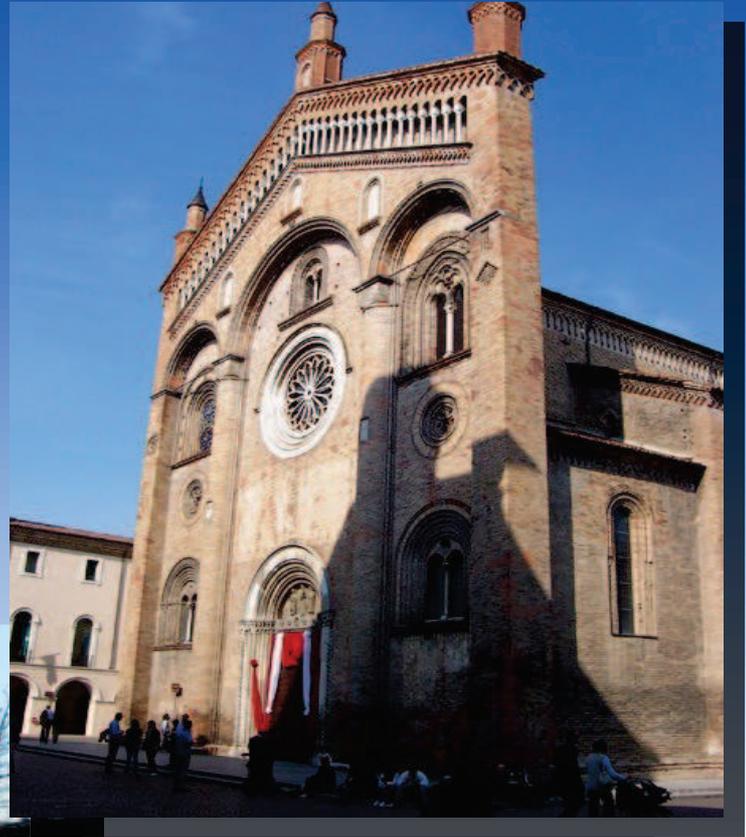
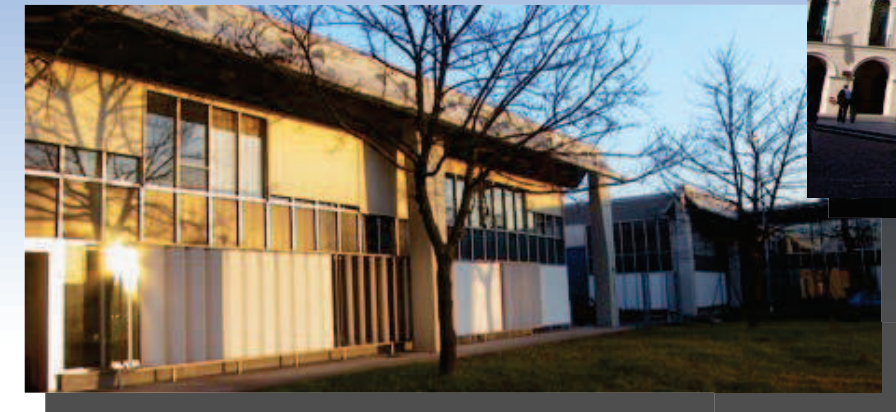
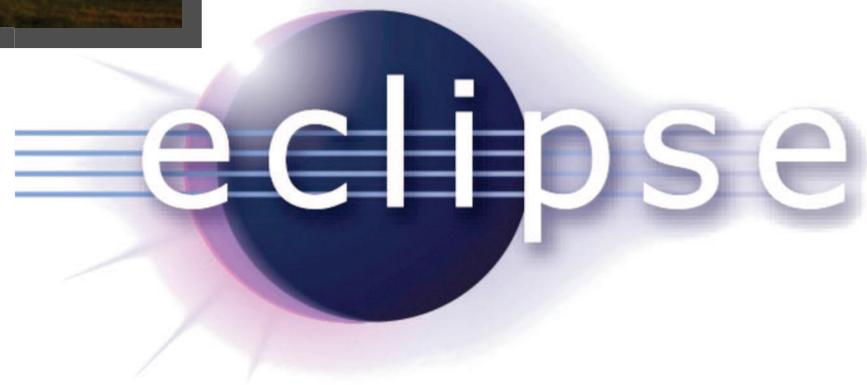

# Eclipse-IT 2013

**VIII Workshop of the Italian Eclipse Community**

**Crema, September 19-20 2013**

**Elvinia Riccobene (ed.)**

Elvinia Riccobene (ed.)

# Eclipse-IT 2013

VIII Italian Workshop on Eclipse Technologies
Crema, September 19-20, 2013
Università degli Studi di Milano

# Preface

This volume contains the extended abstracts of the contributions presented at EclipseIT 2013, the 8[th] workshop of the Italian Eclipse Community, hosted by the Computer Science Department of the University of Milan (Crema Campus) on September 19-20, 2013.

Previous editions took place is Rome (2006), Naples (2007), Bari (2008), Bergamo (2009), Savona (2010), Milan (2011), and Pozzuoli (2012).

Although Eclipse was initially designed as an integrated development environment (IDE) for object-oriented application development, today it represents an open development platform comprised of extensible frameworks, tools and runtimes for building, deploying and managing software.

Around Eclipse, an international live community continuously works on improving the framework and on promoting the use of Eclipse. That happens also in Italy. This workshop is, indeed, the eighth yearly meeting of the Italian Eclipse Community which includes universities, public institutions and industries, researchers and practitioners, students and professionals, all joined by the interest in experimenting, extending, and supporting the Eclipse platform.

The special topic of this edition is the S*oftware cooperative development for mobile applications*. Two tutorials are offered on this theme: (1) *Sviluppo di applicazioni enterprise per il mobile con IBM Jazz, Eclipse e Worklight* by Ferdinando Gorga from IBM, and (2) *Uso di Eclipse per lo sviluppo cooperativo del software*, by Paolo Maresca of the University of Naple, Federico II.

We received six contributions from universities, five from industries, some in collaboration with the academy, and six projects were submitted by students who participate to the selection of the best student project awarded by Seen Solution.

The conference program included also Ralph Muller, Director Eco Systems Europe of the Eclipse Foundation, as keynote speaker. He talks about *new and existing things at Eclipse*. This year, a live streaming of the workshop is available to give the opportunity to a large number of students and professionals to follow the event at a distance.

A further novelty of the current edition is a panel on the theme *Professional skills of ICT professions in Italy on the basis of the European eCompetence Framework (e-CF)*, organized by Domenico Squillace, IBM Italy.

This year, the conference received the patronage of ANORC, the official Italian digital storage responsible association.

Many people contributed to the organization and realization of this workshop. I would like to thank all the participants who contributed to the success of a new edition of EclipseIT. I thank the members of the Program Committee for their evaluable suggestions and the entire Eclipse Italian Community chaired by Paolo Maresca who gave a special contribution for supporting this workshop. I thank Domenico Squillace, who makes possible the realization of a panel on a theme that is extremely important today both for the industry and for the academy. I am grateful for their invaluable work to Paolo Arcaini who chaired the student track, Davide Rebeccani for the technical support, and Claudia Piana for her administrative help. I am thankful to the sponsors of this event, IBM Italy with Carla Milani, the Department of Computer Science of the University of Milan, the Eclipse Foundation, and Seen Solution. Finally, a special immeasurable thank to Angelo Gargantini who helped me in organizing this year workshop and preparing this volume.

Crema, September 2013

Elvinia Riccobene

# Conference Chair

Elvinia Riccobene, Università degli Studi di Milano, Italy

# Industry track Chair

Domenico Squillace, IBM, Italy

# Program Committee

Francesca Arcelli, Università di Milano Bicocca, Italy

Mauro Coccoli, Università di Genova, Italy

Andrea De Lucia, Università di Salerno, Italy

Giovanni Denaro, Università di Milano Bicocca, Italy

Giacomo Franco, IBM, Italy

Angelo Gargantini, Università di Bergamo, Italy

Ferdinando Gorga, IBM, Italy

Paolo Maresca, Università di Napoli Federico II, Italy

Antonio Natali, Università di Bologna, Italy

Patrizia Scandurra, Università di Bergamo, Italy

Lidia Stanganelli, Università di Napoli, Italy

Gianni Vercelli, Università di Genova, Italy

# Student track Chair

Paolo Arcaini, CNR IDPA Dalmine, Italy

# Technical support

Davide Rebeccani. Università degli Studi di Milano, Italy

# Sponsors

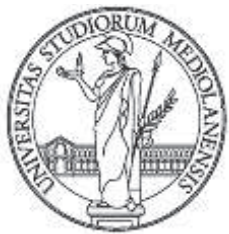 Università degli Studi di Milano

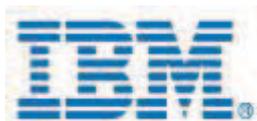 IBM

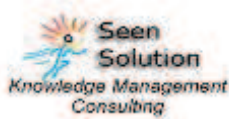 Seen Solution

# Patronage

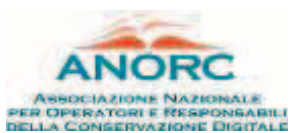 ANORC

# Sommario

## Research track

## Student track

# Parsley for your EMF Applications
# (Extended Abstract)

Lorenzo Bettini
Dipartimento di Informatica,
Università di Torino, Italy
bettini@di.unito.it

Vincenzo Caselli
RCP Vision
vincenzo.caselli@rcp-vision.com

Francesco Guidieri
RCP Vision
francesco.guidieri@rcp-vision.com

EMF, together with JFace, databinding and persistence implementations, simplifies the development of complex software applications with its modeling mechanisms. However, setting up and assembling all these technologies still requires both deep knowledge of most internal details and a considerable amount of time; the development of applications based on EMF could be made simpler, with more code reuse, and without having to deal with too many internal details.

In this talk we present EMF Parsley, a framework that has been recently approved as an Eclipse project. It is based on the previous experimental framework EMF Components [4].

EMF Parsley provides a framework to easily develop UI components based on EMF models. The framework hides most of the complexity of internal details: creating a JFace viewer and connecting it to an EMF resource, usually requires a few lines of code. The main design principle underlying this framework is to split responsibilities into small classes (adhering to the Single Responsibility Principle [9]); thus, customizing a single aspect of the components requires to specialize only the class that deals with that specific aspect, not the whole component. This should maximize code reuse and promote a programming style where the classes implemented by the programmer are usually very small and deal with not too many aspects.

The framework also comes with some UI components that can be used out-of-the-box (including trees, tables and forms, and view and editor parts). Programmers can rely on these components and customize them, or use them as a reference implementation to build their own components based on our framework.

To customize existing components we rely on Dependency Injection [7], in particular using Google Guice. The configuration and setup of Guice modules uses the same mechanism of Xtext (i.e., binding an implementation class only requires to write a specific "bind" method in the main Guice module class). The initial setup of a project which uses EMF Parsley is done via a project wizard, so that the programmer does not have to deal with these details.

Specification of custom behaviors (e.g., label providers, content providers, context menus, etc.) is based on the types of elements of the EMF model, but without requiring long cascades of Java "instanceof" and casts: we provide a polymorphic method dispatch mechanism (borrowed from Xtext) that allows to write cleaner and declarative code.

We also provide a DSL (implemented in Xtext [3, 5]) for making the use of our framework easier: customizations can be specified in a compact form in a single file. The DSL compiler will automatically generate all the Java classes and the corresponding Guice module bindings. The DSL provides a fully feature Eclipse editor and, by relying on Xbase [6], it provides a Java-like language completely integrated with Java and Eclipse JDT.

Setting up views (like in Figure 1) only requires a few lines of Java code. Using the DSL (like in Figure 2) the customization of the components is really quick (the corresponding Java code will be automatically generated by the DSL compiler).

Concerning the persistence aspect, the framework can handle a generic EMF persistence implementation, like XMI or Teneo, just providing the EMF Resource's URIs. It also includes a bundle for handling CDO resources, which takes care of CDO sessions and transactions transparently.

All sources are covered by tests, both unit tests (with JUnit) and functional tests (with SWTBot). Tests are already integrated in the Continuous Integration system implemented with Jenkins.

Also the building and update site architecture is already setup and implemented with Buckminster [8]. This building tool is heavily used throughout the whole development cycle, starting from the workspace and platform materialization up to headless building and continuous integration.

In addition to be used to render RCP UI, the framework is out-of-the-box ready to be used for RAP development (Remote Application Platform [2]). By using single sourcing techniques, it is easy to develop an RCP application with EMF Parsley and to automatically have a web version (based on RAP) of the same application. EMF Parsley can also be used with the new Eclipse e4 platform.

The framework that is closer to our proposal is EMF Client Platform [1]. However, while the latter aims at providing a quick way of obtaining a full application based on EMF, EMF Parsley aims at providing single smaller and reusable components. Moreover, our components are not customized via extension points, but using injection with plain Java and with a DSL.

The framework presented here is available at http://www.eclipse.org/emf-parsley.

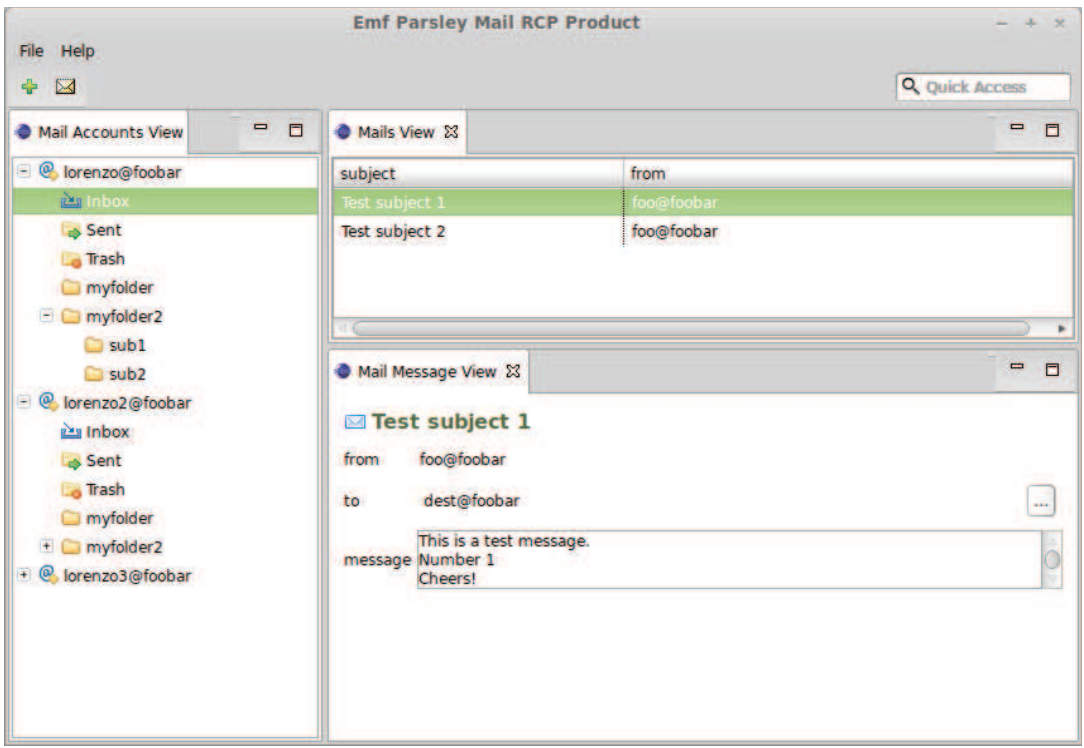

Figure 1: A Mail RCP Example implemented with EMF Parsley.

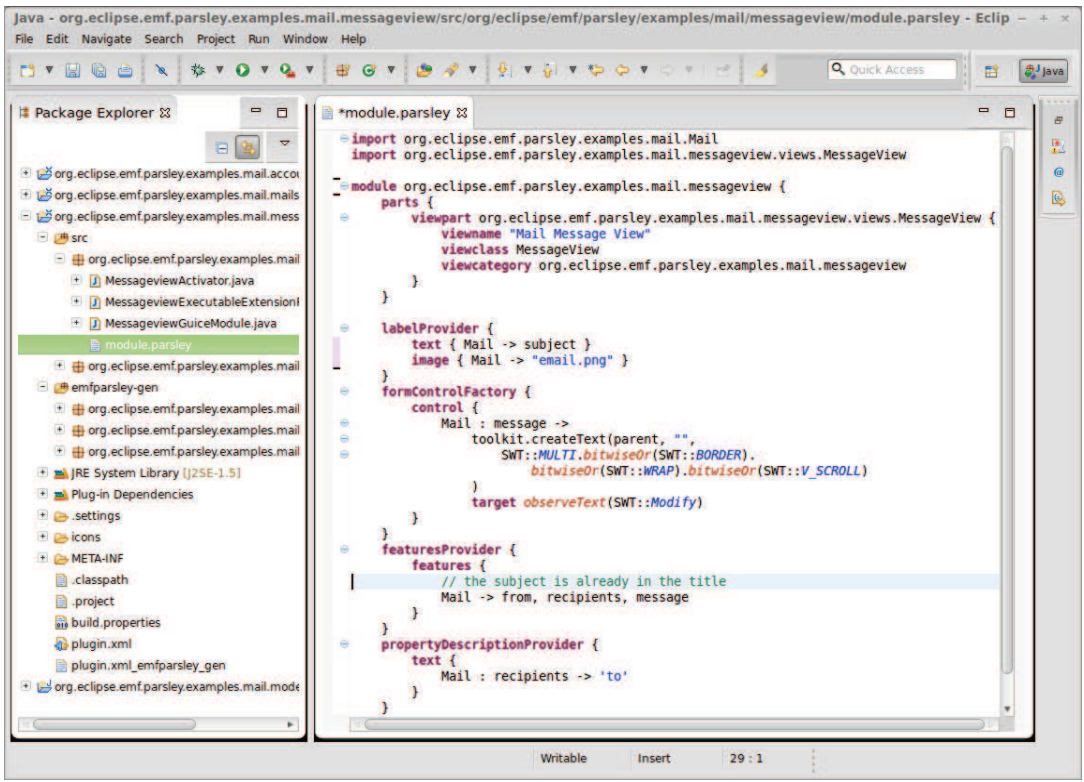

Figure 2: Using the EMF Parsley DSL for customizations.

# A Java-Based Certification Framework for Service Compositions

Marco Anisetti, Claudio A. Ardagna, Ernesto Damiani

Dipartimento di Informatica
Università degli Studi di Milano
Crema (CR), 26013, Italy
firstname.lastname@unimi.it

Federico Giuba

Università degli Studi di Milano
Crema (CR), 26013, Italy
federico.giuba@studenti.unimi.it

## ABSTRACT

The evaluation of security properties of web services is a key problem, especially when business processes are dynamically built by composing atomic services provided by different suppliers at runtime. In this paper we present a Java-based framework for the security certification of BPEL-based compositions of web services. The framework is grounded on the test-based service certification scheme proposed in the context of the ASSERT4SOA Project (http://www.assert4soa.eu/) and permits to virtually certify a BPEL-based composite service for a set of security properties, starting from certificates awarded to the component services.

## 1. INTRODUCTION

The success of web services is deeply changing the design, development, and distribution of software. Today, business processes are often implemented through a dynamic composition of web services available over the Internet. In a context where web services are continuously (re-)designed and released, an accurate and robust evaluation of composite service security becomes a fundamental challenge. The use of assurance solutions based on certification to evaluate software security has recently received considerable attention [4], but most of the available solutions have not considered the certification of composite services or are not suitable for dynamic run-time compositions.

Recently, Anisetti et al. [1] have proposed a test-based security certification scheme for BPEL-based composite services, where the basic idea is to produce a virtual security certificate for compositions on the basis of the certificates of component services. Here, the term "virtual" refers to the fact that the test evidence in the certificate proving a given security property for the composition is inferred by the certification authority, with no real testing activities. In this paper we describe a Java-based framework (developed using Eclipse IDE) implementing the approach in [1] and present a preliminary performance analysis.

## 2. BPEL SECURITY CERTIFICATION FRAMEWORK

Our certification approach models a BPEL process as a *BPEL graph* in which each web service invocation is represented as a vertex. Service invocations are connected together following different compositional pattern: sequence, alternative, or parallel. An *annotated BPEL graph* extends a BPEL graph with a labeling function, which annotates every vertex representing an invoke operation with functional and security requirements. Given an annotated BPEL graph, a *BPEL instance graph* is defined, where every vertex representing an invoke operation is instantiated with a real service such that the following conditions hold: *i)* the service satisfies the functional requirements and *ii)* its certificate satisfies security requirements in the annotations. Our BPEL security certification framework receives as input an annotated BPEL graph and a set of candidate services with the corresponding security certificates (following the certification scheme described in [1]), selects the best subset of services satisfying the BPEL security annotations (BPEL instance graph), and generates the virtual security certificate for the BPEL instance graph as output. Figure 2 shows an architectural overview of the BPEL security certification framework.

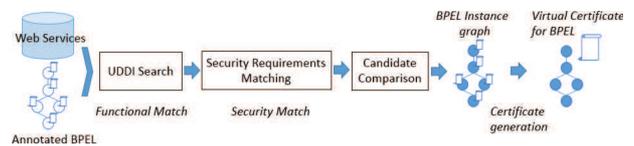

**Figure 1: Architectural overview of the BPEL security certification framework**

The BPEL instance graph is generated adopting a four-step selection process as follows: *i) UDDI Search*, for each invoke operation in the annotated BPEL graph the framework performs an *UDDI inquiry* obtaining a set of candidate services that satisfy functional requirements; *ii) Security Requirements Matching Process*, the security certificates of every candidate service selected at step *i)* are matched with security requirements defined in the annotated BPEL graph; *iii) Comparison Process*, candidates returned at step *ii)* are ranked on the basis of their security properties; *iv) BPEL instantiation*, the first candidate service in the ranked list of step *iii)* is associated to the BPEL instance graph invocation as a partner link. As soon as the BPEL instance graph is generated, a *Virtual Certification Process* is triggered. This process evaluates *i)* the security certificates of the candidate services belonging to the BPEL instance graph and *ii)* the compositional pattern defined by the BPEL graph. It then generates the virtual certificate for the BPEL instance. In the following we describe our components in detail.

```
13:50:32.827 INFO [Invoke] : MATCHING WS candidates for Partnerlink=ns2:SecureStorage

***************** MATCHING TEST *****************

          ----------ASSERT4SOA-CERTIFICATE----------
          || PROPERTY NAME: org.property.sec.property_13
          || absClass: Confidentiality
          || algo: AES
          || key: 256
          || ctx: in_storage
          ------------------------------------------
    WITH:
          ----------ASSERT4SOA-ANNOTATION-----------
          || absClass: Confidentiality
          || ctx: in_storage
          || sf: encryption
          ------------------------------------------

          RESULT = true

**************************************************
```

Figure 2: An example of matching

## 2.1 UDDI Component

For each invoke in the annotated BPEL, a set of services compatible with the functional annotations are selected. In particular, this selection is implemented via inquiring a UDDI registry (method `findCandidates()` of class *BpelGraph*). Our framework integrates the Apache jUDDI-client [3], based on UDDI v3 API, that allows to connect to every UDDI v3 compliant registry. The functional service search is performed using tModels [5] in the UDDI structure. Searching for services compatible with a specific tModel, we obtain a set of endpoints identifying partner link candidates that satisfy functional requirements.

## 2.2 Matching Process Component

It receives as input, from the UDDI Component, a set of web services compatible with the functional requirements expressed for each invocation in the annotated BPEL. This component implements the matching algorithm (method `matchCandidates()`) described in [2] allowing to select, between the functional compatible services, those candidate services that satisfy the security requirements expressed in the annotated BPEL. Figure 2.2 shows an example of matching process between information stored in a certificate and the requirements expressed in the annotation.

## 2.3 Comparison Process Component

It receives as input, from the Matching Process Component, a set of functional and security compatible candidate services for every BPEL invocation. This component implements the comparison algorithm described in [2], ranking candidate services based on their security certificates. This component exposes method `compareCandidates()` that receives as input a set of partner link candidates and sorts them out on the basis of their certificates, according to the "property first" approach described in [2].

## 2.4 Virtual Certification Component

It receives as input a BPEL instance graph and produces as output a virtual certificate for it. To this aim, it iteratively composes pairs of (virtual) certificates $C_i$ and $C_j$, according to their compositional pattern, to generate a virtual certificate for the composition. The process is repeated until the BPEL instance graph is reduced to a graph having a single vertex $v$ with virtual certificate $C_{ij}^*$. Given two certificates $C_i$ and $C_j$, a virtual certificate $C_{ij}^*$ is generated including a security property, a model of the service, and the evidence supporting the security property. At the

```
COMPOSING PARALLEL invokes: ID=3 with ID=5

--BpelGraph status:
BpelVertex [ID=0, parent=-1, type=sequence, children=(1, 6, 11, ), name=main]
BpelVertex [ID=1, parent=0, type=parallel, children=(3, ), name=Flow]
BpelVertex [ID=3, parent=1, type=invoke, children=(), name=3+5]
BpelVertex [ID=6, parent=0, type=alternative, children=(8, 10, ), name=Airline_BookFlight]
BpelVertex [ID=8, parent=6, type=invoke, children=(), name=Invoke_BookFlight_A]
BpelVertex [ID=10, parent=6, type=invoke, children=(), name=Invoke_BookFlight_B]
BpelVertex [ID=11, parent=0, type=invoke, children=(), name=11+12]

Certificate of ID=3
          ----------ASSERT4SOA-CERTIFICATE----------
          || PROPERTY NAME: org.property.sec.property_10
          || absClass: Confidentiality
          || algo: AES
          || key: 128
          || ctx: in_storage
          ------------------------------------------

Certificate of ID=5
          ----------ASSERT4SOA-CERTIFICATE----------
          || PROPERTY NAME: org.property.sec.property_13
          || absClass: Confidentiality
          || algo: AES
          || key: 256
          || ctx: in_storage
          ------------------------------------------

Generated virtual certificate
          ----------ASSERT4SOA-CERTIFICATE----------
          || PROPERTY NAME: Virtual Certificate Property
          || absClass: Confidentiality
          || algo: AES
          || ctx: in_storage
          || key: 128
          ------------------------------------------
```

Figure 3: An example of service composition. First, the details about the BPEL graph are shown; then, service certificates to be composed are listed; finally, the virtual certificate is presented

Table 1: Performance analysis.

| Number of candidate services per invoke | Ad-hoc rules | Execution times (s) |
|---|---|---|
| 10 | 20 | 1.28s |
| 50 | 20 | 1.90s |
| 200 | 20 | 3.78s |
| 10 | 100 | 5.55s |
| 50 | 100 | 6.05s |
| 200 | 100 | 7.76s |

current stage our framework supports a virtual certificate composition process working at security property level only.[1] In particular, a virtual property can be generated either by applying a set of *ad-hoc* rules defined by experts, which specifies if and how a virtual security property can be generated from the composition of two different security properties, or with a *default* rule, which considers relationships among properties defined in a property hierarchy. Figure 2.4 shows a composition of a pair of certificates in parallel.

## 3. PERFORMANCE ANALYSIS

Performance analysis has been executed on a workstation equipped with Intel Core i5-2500K 3.30GHz, 8 GB of RAM, 128 GB of SSD, running Windows 7 64-bit and Java 1.7.0-21, and has considered a BPEL process with 6 invoke elements.

Table 1 shows a detailed performance evaluation by increasing the number of candidate services per invoke and the number of ad-hoc rules to be evaluated for the generation of the virtual security property. Our results show that the time needed for a runtime generation of a virtual certificate is reasonable and is 7.76s in the worst-case scenario with 100 ad-hoc rules and 200 candidate services for each invocation.

---

[1] We leave the support for virtual models and evidence to our future work.

## 4. CONCLUSIONS

We presented a Java-based framework (developed using Eclipse IDE) for the security certification of BPEL-based compositions. The proposed framework generates a virtual test-based certificate for a composite service, which consists of a virtual security property, starting from the certificates of the component services. Our future work will consider the implementation of an extended framework providing a solution for the generation of complete virtual certificates.

# ARIES: An Eclipse plug-in to Support Extract Class Refactoring

Gabriele Bavota
University of Sannio
gbavota@unisannio.it

Andrea De Lucia
University of Salerno
adelucia@unisa.it

Andrian Marcus
Wayne State University
amarcus@wayne.edu

Rocco Oliveto
University of Molise
rocco.oliveto@unimol.it

Fabio Palomba
University of Salerno
fabio.palomba.89@gmail.com

Michele Tufano
University of Salerno
tufanomichele89@gmail.com

## ABSTRACT

During Object-Oriented development, developers try to define classes having (i) strongly related responsibilities, i.e., high cohesion, and (ii) limited number of dependencies with other classes, i.e., low coupling [1]. Unfortunately, due to strict deadlines, programmers do not always have sufficient time to make sure that the resulting source code conforms to such a development laws [8].

In particular, during software evolution the internal structure of the system undergoes continuous modifications that makes the source code more complex and drifts away from its original design. Classes grow rapidly because programmers often add a responsibility to a class thinking that it is not required to include it in a separate class. However, when the added responsibility grows and breeds, the class becomes too complex and its quality deteriorates [8]. A class having more than one responsibility has generally low cohesion and high coupling. Several empirical studies provided evidence that high levels of coupling and lack of cohesion are generally associated with lower productivity, greater rework, and more significant design efforts for developers [6], [10], [11], [12], [13]. In addition, classes with lower cohesion and/or higher coupling have been shown to correlate with higher defect rates [9], [14], [15].

Classes with unrelated methods often need to be restructured by distributing some of their responsibilities to new classes, thus reducing their complexity and improving their cohesion. The research domain that addresses this problem is referred to as refactoring [8]. In particular, *Extract Class* Refactoring allows to split classes with many responsibilities into different classes. Moreover, it is a widely used technique to address the Blob antipattern [8], namely a large and complex class, with generally low cohesion, that centralize the behavior of a portion of a system and only use other classes as data holders. It is worth noting that performing Extract Class Refactoring operations manually might be very difficult, due to the high complexity of some Blobs. For this reason, several approaches and tools have been proposed to support this kind of refactoring. Bavota et. al [2] proposed an approach based on graph theory that is able to split a class with low cohesion into two classes having a higher cohesion, using a MaxFlow-MinCut algorithm. An important limitation of this approach is that often classes need to be split in more than two classes. Such a problem can be mitigated using partitioning or hierarchical clustering algorithms. However, such algorithms suffer of important limitations as well. The former requires as input the number of clusters, i.e., the number of classes to be extracted, while the latter requires the definition of a threshold to cut the dendogram. Unfortunately, no heuristics have been derived to suggest good default values for all these parameters. Indeed, in the tool JDeodorant [7], which uses a hierarchical clustering algorithm to support Extract Class Refactoring, the authors tried to mitigate such an issue by proposing different refactoring opportunities that can be obtained using various thresholds to cut the dendogram. However, such an approach requires an additional effort by the software engineer who has to analyze different solutions in order to identify the one that provides the most adequate division of responsibilities.

We tried to mitigated such deficiencies by defining an approach able to suggest a suitable decomposition of the original class by also identifying the appropriate number of classes to extract [3, 4]. Given a class to be refactored, the approach calculates a measure of cohesion between all the possible pairs of methods in the class. Such a measure captures relationships between methods that impact class cohesion (e.g., attribute references, method calls, and semantic content). Then, a weighted graph is built where each node represents a method and the weight of an edge that connects two nodes is given by the cohesion of the two methods. The higher the cohesion between two methods the higher the likelihood that the methods should be in the same class. Thus, a cohesion threshold is applied to cut all the edges having cohesion lower than the threshold in order to reduce spurious relationships between methods. The approach defines chains of strongly related methods exploiting the transitive closure of the filtered graph. The extracted chains are then refined by merging trivial chains (i.e., chains with few methods) with non trivial chains. Exploiting the extracted chains of methods it is possible to create new classes - one for each chain - having higher cohesion than the original class.

In this paper, we present the implementation of the proposed Extract Class Refactoring method in ARIES (Automated Refactoring In EclipSe) [5], a plug-in to support refactoring operations in Eclipse. ARIES provides support for Extract Class Refactoring through a three steps wizard.

In the first step, shown in figure 1, the tool supports the software engineer in the identification of candidate Blobs through the computing of three quality metrics, namely LCOM5 [6], C3 [9] and MPC [16]. Thus, ARIES does not compute an overall quality of the classes, but it considers only cohesion and coupling as the main indicators of class quality in this context. Hence, Blobs are usually outliers or classes having a quality much lower than the average quality of the system under analysis [9]. The identification of Blobs in ARIES is based on such a conjecture. In the second step of the wizard, the software engineer has the possibility to further analyze a candidate Blob and get insights on the different responsibilities implemented by analyzing its topic map, represented as the five most frequent terms in a class (the terms present in the highest number of methods). For this reason, the topic map is represented by a pentagon where each vertex represents one of the main topics. Once a class that needs to be refactored is identified, the software engineer activates the last step of the wizard (shown in figure 2) to obtain a possible restructuring of the class under analysis. ARIES reports for each class that should be extracted from the Blob the following information: (i) its topic map; (ii) the set of methods composing it; and (ii) a text field where the developer can assign a name to the class. The tool also allows the developer to customize the proposed refactoring moving the methods between the extracted classes.

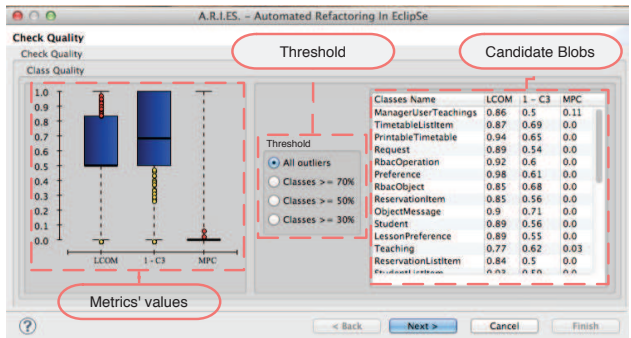

Figure 1: ARIES: Identification of candidate Blobs.

In addition, ARIES offers the software engineer on-demand analysis of the quality improvement obtained by refactoring the Blob, by comparing various measures of the new classes with the measures of the Blob. When the developer ends the analysis, the extraction process begins. ARIES will generate the new classes making sure that the changes made by the refactoring do not introduce any syntactic error. A video of the tool is available on Youtube[1].

[1] http://www.youtube.com/watch?v=csfNhgJlhH8

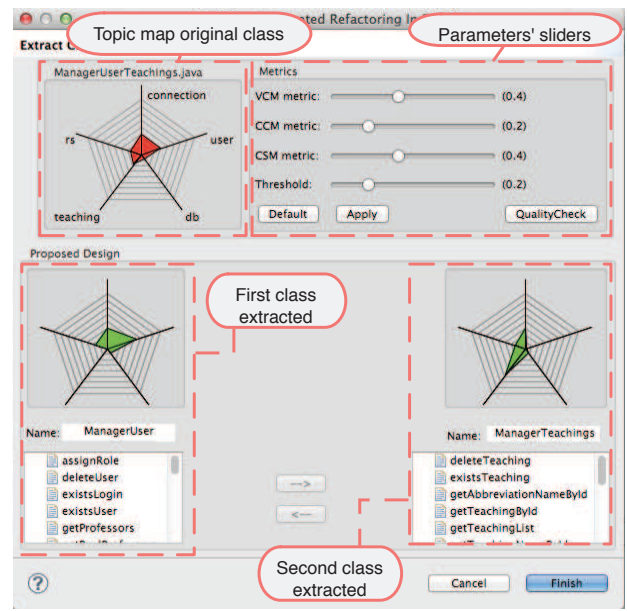

Figure 2: ARIES: Extract Class refactoring.

# An Eclipse framework to ease notification among *MyUniBG* app[*]

Steven Capelli
Università degli Studi di Bergamo
DIIMM
Dalmine, Italy
steven.capelli@unibg.it

Andrea Ghetti
Università degli Studi di Bergamo
DIIMM
Dalmine, Italy
andrea.ghetti@studenti.unibg.it

Davide Mora
Università degli Studi di Bergamo
DIIMM
Dalmine, Italy
davide.mora@unibg.it

Simone Mutti
Università degli Studi di Bergamo
DIIMM
Dalmine, Italy
simone.mutti@unibg.it

## ABSTRACT

The widespread diffusion of Android OS has led to a rapid explosion of the Google Play store (previously Android Market). As of 2011, the Play store includes more applications than the Apple App Store. It is natural to compare this growth to what happened years ago in the area of the World Wide Web, although in the second case, the need was to allow users to interact and collaborate in a more easy way (e.g., social network), while now the need is to try to use the same device everywhere (e.g., office, home) and for everything (e.g., work, free time). This reflects the rising concept of "Bring Your Own Device" (BYOD). In this vision, University of Bergamo decided to develop its own Android application (named MyUniBG) in order to provide to their students and staff members advanced features (e.g., information about lessons hours and course details). In this paper, we want to highlight how, thanks to the support provided by Eclipse framework, we can make a step toward the "BYOD vision".

The aim of the tool is to ease the notification (e.g., variations in lessons hours) using Google Cloud Messaging for Android, a service that allows you to send data from your server to your users' Android-powered device, and also to receive messages from devices on the same connection. The tool, implemented as an Eclipse RAP plug-in, will provide two main functionalities: (a) *"push"* information to the community of devices (communication server-client), and (b) *"pop"* notifications from a device and share this information with others. In this scenario, MyUniBG app will be extended in order to be the front-end for the new features.

## Keywords
RAP, Android app, Google Cloud Messaging

## 1. INTRODUCTION

[*]This work was partially supported by the EC within the 7FP, under grant agreement 257129 "PoSecCo", by the Italian Ministry of Research within the PRIN projects "PEPPER 2008", "GATECOM" and "GenData 2020".

The widespread diffusion of Android OS has led to a rapid explosion of the Google Play store (previously Android Market). As of 2011, the Play store includes more applications than the Apple App Store. At press time Android has reached in Europe around the 70% of the market share of smartphones [4]. It is natural to compare this growth to what happened years ago in the area of the World Wide Web, although in the second case, the need was to allow users to interact and collaborate in a more easy way (e.g., social network), while now the need is to try to use the same device everywhere (e.g., office, home) and for everything (e.g., work, free time). This reflects the rising concept of "Bring Your Own Device" ($BYOD$). In this vision, University of Bergamo decided to develop its own Android application (named *MyUniBG*) in order to provide to their students and staff members advanced features (e.g., information about lessons hours and course details). The previous version of MyUniBG offers some minor functionalities: daily schedule and timetables, staff list with contacts, shortcuts to the main website (*www.unibg.it*) and access to university email *@studenti.unibg.it* .

Before our tool, each variations or information regarding students could be viewed only thought the website in a non-user-friendly way: students should visit everyday that page in order to know if a lesson has been suspended or delayed. As we have written above the Android diffusion permits to take advantage of this scenario allowing the developers to create a synergy between services provided by University, students own devices and Eclipse framework. This situation also enforces the BYOD concept: each student, with his own device, could benefit of all this services without buying or using other tools.

The aim of the tool is to ease the notification (e.g., variations in lessons hours and notifications of various information) to the end users besides the functionalities already provided by the app. This was made possible using Eclipse RAP and Google Cloud Messaging for Android, a communication protocol offered by the official Android APIs that allows you to send data from your server to your users' Android-powered device, and also to receive messages from devices on the same connection.

The tool provides two main functionalities (a) *"push"* information to the community of devices (communication server-client), and (b) *"pop"* notifications from a device and share this information with others. The server application has been built upon Eclipse Remote Application Platform (RAP) and manages the exchanged data between server-devices and device-server. The client device receives a small amount of data and shows the user the info transmitted.

The paper is organized as follows. Section 2 presents the main features provided by the Eclipse RAP application. Section 3 briefly illustrates the Google Cloud Messaging protocol. Section 4 and 5 discuss, respectively the design principles followed in the design of the Server and Client parties. Finally, Section 6 draws some concluding remarks.

## 2. ECLIPSE RAP SERVER APPLICATION

*Remote Application Platform* [8] (RAP, formerly Rich Ajax Platform) Project is an open-source software project under the Eclipse Technology Project which aims to enable software developers to build Ajax-enabled rich Internet applications by using the Eclipse development model, plugins and a Java-only application programming interface (API). The goal of RAP is to move RCP applications with minimal effort into a web browser. Thus, they can be used from everywhere over the web without the need of a local installation. A standard web browser is sufficient. RAP is very similar to Eclipse RCP, but instead of being executed on a desktop computer, RAP applications run on a server and standard browsers are used on the client side to display the GUI. One central part of RAP is AJAX (Asynchronous JavaScript and XML). A browser can communicate with a server via AJAX requests. This allows changing small parts of a web page without the need to reload it completely. With this ability, it is possible to build complete applications that seem to be executed within a browser. To be precise, this means that the major part of the application runs on the web server. The data structures are stored, accessed and modified on the server. Furthermore, the server controls the logic of the user interface on the client. The client has the look and feel of an application, but it displays only the GUI and it renders the data it receives from the server.

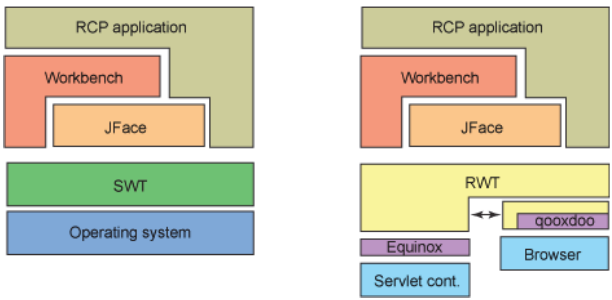

**Figure 1: Remote Application Platform**

According to the common problems that occur in a big structure, there are more than a few people that should be able to modify or update the daily schedule for lessons, e.g. administrative employees, professors or ushers. Each one should be able to do this task and a web interface looks like the better approach to this situation. RAP is ideal for this scenario, as the RAP project enables developers to build rich, AJAX-enabled web applications by using the Eclipse development model, plug-ins with the well known Eclipse workbench extension points, Jface, and a widget toolkit with SWT API. Developers of a web application implement the GUI with the Java interface of the SWT (Standard Widget Toolkit) as they would do for an Eclipse application. In RAP the implementation of SWT is replaced by RWT, which uses Qooxdoo for the client-side GUI presentation. Qooxdoo [7] is an Open Source project that aims at creating application-like GUIs in web browsers with the help of JavaScript and Ajax [5, 6]. The backend, i.e. the access to the data structures in Java, does not have to be changed at all. Java objects do not have to be translated into streams of Ajax requests. RAP takes advantage of the Java development tools and the plug-in development tools provided by Eclipse. The applications are entirely developed in Java as bundles (plug-ins). The services, from development to launching, debugging and exporting to standard .war files, work out of the Eclipse IDE (a Web archive (WAR) file is a packaged web application). RAP enables application developers familiar with RCP to create web application using the RCP programming model. They do not need to be experts about JavaScript and server-side scripting languages.

However, there is one important difference between RCP and RAP applications. In most cases an RCP application performs operations on its own data set. For instance, a user opens his own files with exclusive write access. In a RAP application a user normally does not have a private, exclusive data set. With a web application, users do usually access the same database. This is not a problem, as long as all users have only read access. Special care has to be taken if the users are allowed to modify their common data. For instance, it might be required to update the data representation in the web application of all users that are logged in, when one user changes something.

The main advantages of RAP in the previously described scenario are:

- The implementation looks like a real application that runs in a browser. It does not have to be installed. This is particularly interesting for the casual user;

- It is platform independent;

- There are benefits for collaborative work, as several users can share the same data, which is located on the server;

- The same Java code base is shared for RCP and RAP applications and the development is completely in Java. This is a significant benefit for the developers and the code quality of the application.

Common disadvantages of web based solutions must be considered:

- Nearly every event in the GUI triggers an Ajax call to the server, e.g., opening a (contextual) menu. Depending on the speed of the network and the responsiveness of the server, the workflow can be slowed down considerably. This might affect the user motivation;

- Working without an internet connection is impossible;

- Slow client machines can suffer performance issues.

## 3. GOOGLE CLOUD MESSAGING

*Google Cloud Messaging for Android* [2] (GCM) is a free service by Google that helps developers sending data from servers to their Android applications on Android devices, and upstreaming messages from the user's device back to the cloud. This message could be an acknowledgment, an update or anything that fits up to 4kb of payload data. The GCM service handles all aspects of queuing of messages and delivering to the target Android application running on the target device [3].

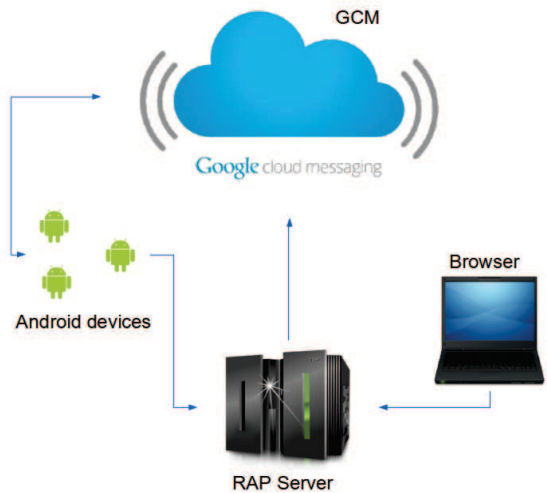

**Figure 2: Relationship between server, app and GCM**

GCM has these main features:

- It allows RAP application to send messages to MyUniBG app;

- Using the GCM Connection Server, the server can receive upstream messages from the user's device;

- MyUniBG app doesn't need to be running to receive messages. The system will wake up the application via Intent broadcast when the message arrives, as long as the application is set up with the proper broadcast receiver and permissions. This behavior contributes to a better smartphone power management;

- It has no hidden cost outside the data connection for both sides.

As stated by Figure 2 there is a connection cloud between server and app: the server sends a message to the cloud, which forwards it to the target devices. Each devices has its own ID: on the first run of the app, the device registers himself to the Google connection cloud and to the server, notifying in this way that it wants to receive messages. The server then stores this ID: when needed, it sends to the cloud the message targeting only specifics IDs. If a user doesn't want to receive any more further messages he simply needs to unregister only from the cloud: he won't receive anything till he register again.

## 4. SERVER ARCHITECTURE

The core of the architecture, showed in Figure 2, is represented by the server (i.e., the RAP application). As described before, it provides the functionalities to (a) register devices and (b) receive and send notification. The motivation underlying the implementation of the RAP application is that Eclipse is the most used open source development framework, offering a high degree of flexibility and supporting the extention of its functionality through the implementation of plug-ins. Moreover, Eclipse can provide a common repository model as well a common plugin interface to facilitate the integration between this and other plugins, which will be used to manage other common functionalities (e.g., save file). Finally, the Eclipse platform encourages the reuse of the functions of other plug-ins and modules of the Eclipse framework, speeding up and improving the quality of the development process. Hence, the RAP application is composed by the following components (see Figure 3):

- *User login* provides to different users to logon in the application. The login grants to the user the capability to enable a role (e.g., staff, administrator) in order to customize the application based on the user;

- *Backend* implements the logic of the RAP application, it permits to save the registration ID into the SQLite database, to receive and to send messages, it provides the set of statistics about the messages received and sent;

- *Web interface* implements the GUI;

- *GCM subscriber* delivers functionalities to authenticate the RAP application to the GCM and communicate with it.

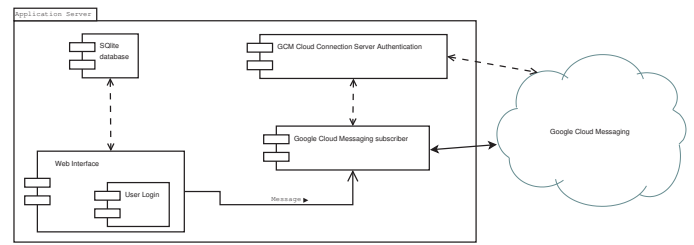

**Figure 3: Application Server Architecture**

All of these components permit the RAP application to send and receive notifications from MyUniBG app. Basically, the app registers the device ID to the GCM and to the RAP application, specifying which news should be aware of. The RAP application stores this data in the SQLite database, then when the timetable needs to be updated, the person in charge sends, through the application frontend, the update, choosing the target of the notification (e.g., lesson of chemistry). The RAP application dispatches to the GCM only to the interested ID and the message to share and finally the app receives the message (silently and without any user update action) and notifies to the user that something has changed in the timetable. If needed a student could reply to the application providing further information about the variation (e.g., "Professor is in a traffic jam").

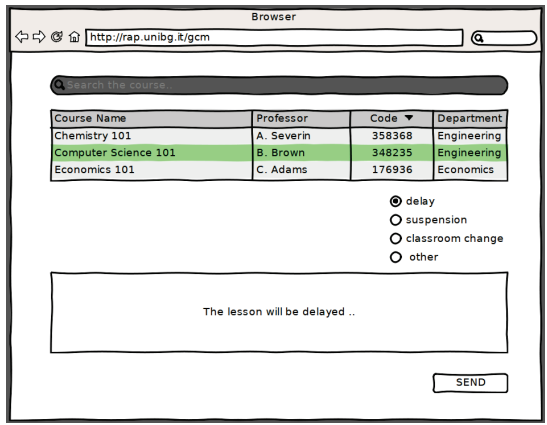

Figure 4: Application Server UI

## 5. CLIENT ARCHITECTURE

As mentioned before our client app was realized as an integration to a previously released app, MyUniBG. This integration has required a little reorganization of the app structure and a partial rewriting of the code. In this way we provided to the end user (and also to future developers) a better user experience and an easier maintainability. MyUniBG provides some functionalities to its users, but only the one related to GCM will be analyzed here. For completeness, however, we say that MyUniBG is based on Android API 18 ( compatible backward to API 8), and interacts with the Unibg servers through XML pages.

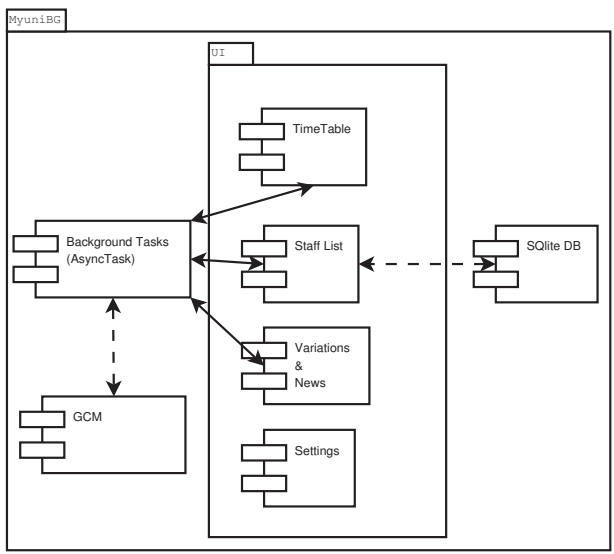

Figure 5: Application Architecture

The GCM is invisible to the end user: this service acts in the background of the UI of the app and notifies only when needed. After the registration phase (explained in the Section 3), this service is automatically woken up by the Android OS. A message (up to 4kb) reports the information encapsulated in a Bundle Object [1]: the app reads this bundle and rearranges each single data in a standard Android notification (see Figures 6 and 7). When the user clicks on the notification the app opens in the variations section, so as to show the variations list.

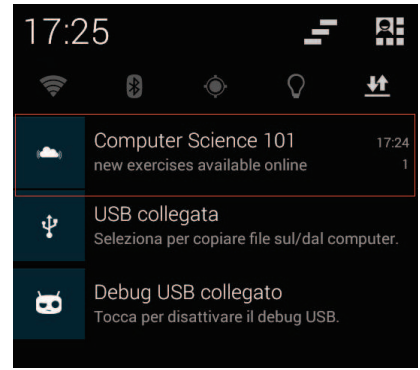

Figure 6: Notification to user

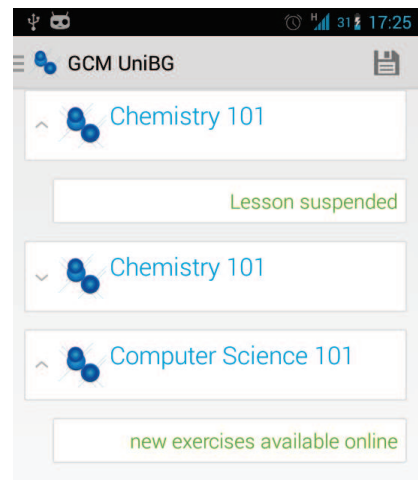

Figure 7: Notifications list in the app

## 6. CONCLUSIONS AND FUTURE WORK

In this paper, we briefly described the structure of web application, implemented by Eclipse RAP framework. We believe that a tool like the one presented in this paper can offer great support in the development of a new way to communicate using common devices (i.e., smartphone). The support provided by the Eclipse framework in the construction of such a system is essential, because Eclipse provides a powerful environment for tool integration and makes already available a large number of functions that are commonly required in the construction of such systems. The interchangeability between RCP and RAP furthermore provides to developers a fast and simple way for building and maintain softwares: like Java (*"write once, run everywhere"*), the Eclipse framework eases the creation of new applications from local (RCP) to remote (RAP) and vice versa.

## 7. REFERENCES

[1] Google Inc. Android API, Bundle. URL: http://developer.android.com/reference/android/os/Bundle.html.

# Eclipse in a Modern Mobile Development Toolchain

Andrea Dal Passo
Lumata Italy
Milano, Italy
andra.dalpasso@lumatagroup.com

Eros Pedrini
Lumata Italy
Milano, Italy
eros.pedrini@lumatagroup.com

## ABSTRACT

*Eclipse* is the basic Java development tool we use in *Lumata*. It's a quite versatile *Integrate Development Editor* (*IDE*) that can be extended via plug-ins to satisfy virtually every specific need. Currently we use it in conjunction with the following state-of-art development tools: *Maven* (for project management), *FindBugs* (for code analysis), *Spring Tool Suite* (to work with Spring Framework), *Jenkins* (for continuous integration), *SonarQube* (for code analysis), *Artifactory* (for component/library management), *Git* (for source management) and *Jira* (for Agile project management). In this paper we describe our complete toolchain and how Eclipse has been integrated within it.

## Keywords
Case studies, Continuous integration, Eclipse, IDE

## 1. INTRODUCTION

*Lumata* was founded in November 2011 with the goal to provide Operators, Brands, and Advertisers with the tools they need to manage their brand's interaction with consumers over mobile: what we call *Mobile Relationship Management*. Lumata is a carve-out from *Business to Business* assets of *Buongiorno*, the leader in *Business to Consumer* entertainment on mobile, and is backed by *Francisco Partners*, a leading technology focused private equity firm. The deal created a global business with more than 450 employees in 18 markets. Lumata offices are located in London (UK), Milano (Italy), Grenoble (France), Madrid (Spain), Moscow (Russia), and Noida (India).

Some of our featured clients are: *TIM* (Italian mobile operator), *Nestlè* (multinational food and beverage company), *Beeline* (brand by *OJSC VimpelCom*, the second-largest telecommunications operator in Russia), and *BMW* (German automobile, motorcycle and engine manufacturing company). For each of these client, a specific solution has been designed: ranging from engaging mobile game applications (e.g., for BMW), to *complex loyalty campaigns* (e.g., Recharge Win reward TIM program). To manage all these different kinds of applications and to improve the quality of our products and the efficiency during the development, we spent a lot of resources to design a toolchain of applications and the flows among them.

In this paper, we will show how Eclipse is one of the building blocks of our development toolchain and how it interacts with the other component. First, we will describe the main components of our toolchain (Section 2). We will then illustrate where Eclipse is used and the plug-ins we adopted (Section 3). Finally we will describe some of the pitfalls you can face (Section 4).

## 2. THE TOOLS WE USE

The first step in our development process involves the *technical specifications*: without considering the specific methodology used (i.e., traditional waterfall or Agile based on Sprints) the technical specifications have to be inserted into *Jira*[1] as *issues* that will be assigned to developers. Jira is developed by *Atlassian*, and even if it is mainly an issue tracking product, it is often used as project management tool. In this step the *product owners* (i.e., who have the knowledge about the final products goals), the technical team leader, and the quality assurance manager work all togethers to write down the technical specifications.

The second step, when needed, is the project/s setup and it consists in preparing one or more *Maven* configuration files. Apache Maven is a build automation tool based on the concept of a *project object model* (*POM*). Maven manages the build, reporting and documentation of a project using an *XML* configuration file called *pom files*. To build a *Mavenized* project, Maven relies on public online library repositories. Considering that we need to support also dependency among our own libraries, we have a private repository based on *Artifactory*. Artifactory provides out-of-the-box all that is needed to set up and run a robust secured repository. Every project has to be deployed into Artifactory: in fact we combine Maven and Artifactory to deploy our projects into different environments[2]: quality assurance (*QA*), staging, and production.

The source code of every project is stored in a *Git* repository. Git is a *distributed version control* and *source code management* (*SCM*) system. We choose Git as principal source code version system because it permits to manage in *simple* way also very complex development dependencies. In fact one of the most important Git feature is the simplicity to branch a project and then merging it back[3].

Our continuous integration system periodically builds all the projects and, in case something goes wrong, immediately alerts the development team in charge. A build can be considered faulty mainly for two reasons: (i) one or more file that don't compile have been pushed on Git, or (ii) some test don't pass. For the continuous integration support, we rely on *Jenkins*: which is an open source tool written in Java.

---

[1] The name Jira is a truncation of *Gojira*, the Japanese name for Godzilla.

[2] An *environment* is a complex system composed of a virtual or physical machine and its operative system.

[3] One of the Git mantra used is *branch early, branch often*.

Every time Jenkins builds a project, the project is analyzed by *SonarQube* (formerly *Sonar*). Sonar is an open platform managing code quality. As such, it covers the 7 axes of code quality: architecture and design, duplication, unit tests, complexity, potential bugs, coding rules, and comments. The Sonar reports are very important during the code review phase.

The last step of our toolchain is QA check. If the QA step (or the code review phase) found bugs or potential issues, they are prioritized and inserted into Jira. In this way the development flow can proceed considering also these issues.

## 3. ANDROID, JAVA, AND PLUG-INS

In this section we describe our mobile and enterprise development environments, mainly focusing on the most important plug-ins.

### 3.1 Android

For *Android* development we tested different solutions. We started working with *Motorola Design Studio*, a customized version of Eclipse configured with Android in mind. When this solution become deprecated, we moved to the official *Eclipse ADT* [3] plug-in. The transition from Motorola Design Studio to Eclipse ADT was very smooth and we were able to use all the knowledge we acquired without problems.

### 3.2 Java

Our most recent enterprise (*J2EE*) projects are developed with *Java 7*: for this reason we use at least Eclipse Indigo (version 3.7.1): which is the first version of Eclipse that full supports Java 7 features. To simplify their development, we use the *Spring* framework, as it provides a comprehensive programming and configuration model for modern Java-based enterprise applications.

### 3.3 Eclipse Plug-ins

The plug-ins we currently use with Eclipse are illustrated in the following.

*Android Development Tools* (ADT): it is a plug-in for Eclipse, designed to give a powerful, integrated environment in which to build Android applications. The main advantage of this solution is that it give the possibility to use the same IDE for both enterprise and mobile development.

*Maven*: to integrate Eclipse with Maven, we use the *m2ec project* [2]. In fact it provides a first-class Apache Maven support in Eclipse, making it easier to edit, for example, Maven's *pom.xml* or run a build from the IDE.

*Spring Tool Suite* [4] (STS): working with Spring, the STS plug-in is a valuable resource. In fact it provides the best Eclipse-powered development environment for building Spring-powered enterprise applications. It supports the last Spring features and also it expands Eclipse with advanced code completion, content-assist, validation, and so on.

*EGit* [1]: even if usually it is better to use Git manually from the command line, EGit is a valid alternative. EGit is an *Eclipse Team Provider* for Git version control system. This plug-in supports all the main functionality of Git (i.e., commit, fetch, push, branch, and merge) in an integrated environment.

*Jira*: Atlassian provides a *connector* for Eclipse based on the popular task-focused *Mylyn* interface. It is used to receive notifications of changes of Jira issues, and also to create, update, comment and manage them.

*FindBugs* [6]: it permits to analyze Java code to find potential bugs. It is based on the concept of *bug patterns* (i.e., a code idiom that is often an error).

*EclEmma* [5]: it is a free Java code coverage tool for Eclipse. It brings code coverage analysis directly into Eclipse workbench without the need to change the code.

It's important to highlight that the last two tools are used by Sonar to perform the same analysis for its reports.

## 4. WHEN THE TOOL IS NOT SUFFICIENT

Even if the Eclipse plug-ins described in Section 3.3 can help a company to create a good development flow, there are some intrinsic problems that you have to warned about.

The first issue you may face is related to Git. Even if it's a powerful tool and is quite simple to use in general, it can drive you crazy if all the team doesn't agree on the source code versioning and branching policies. Another problem is related to its configuration: all the team must share the same Git configuration, otherwise you may have false conflicts signaled during the source code merging phases.

Maven usually doesn't create any issue, but you can face someone within m2ec plug-in. Such as the following two.

- You cannot use the m2ec plug-in if your Maven project rely on automatic code generation during the project creation. Usually m2ec generates the code correctly, but it's not able to associate the generated code as dependency into the Eclipse project.

- By default the m2ec plug-in enables the automatic workspace projects resolution; this means that if one project depends on another project opened in your workspace, m2ec uses this for dependency resolution. This is useful during the development, but it can create problems when you push your changes on the remote repository. If you forgot to push both the projects you will break the continuous integration process.

## 5. CONCLUSIONS

In this paper, we described the tools we use in Lumata and their interaction with Eclipse. We focused mainly on some of the most important Eclipse plug-ins that can simplify developers lives. We then depicted some of the most common issues that a developer may face.

# Native versus Cross-platform frameworks for mobile application development

Rosario Madaudo
MOVIA SpA, Milano, Italy
rmadaudo@movia.biz

Patrizia Scandurra
Università degli Studi di Bergamo, Dipartimento
di Ingegneria, Dalmine (BG), Italy
patrizia.scandurra@unibg.it

## ABSTRACT

Mobile application development is vibrant and reach of opportunities. However, new development questions arise, including what devices to target and which development frameworks to use for creating simple and reliable applications.

Based on our recent experience on developing for mobile devices, in this paper we compare the two main approaches for developing mobile applications (or simply apps), namely i) by using native APIs of the target platform and ii) by adopting cross-platform environments (mostly WEB-based). In particular, we examine the role of Eclipse in both approaches and present our position on combining preciseness and efficiency of native APIs with the flexibility and automation of the Web-based frameworks to achieve significant boosts in both productivity and quality in mobile application development.

## 1. INTRODUCTION

In recent years, our increasing dependence on mobile applications running on a wide spectrum of new devices, from smartphones to tablets, are rapidly posing new development challenges [1]. With mobile and the proliferation of operating systems – the so called "platforms" (including Android, Apple iOS, Microsoft Windows Mobile and Microsoft Phone 7, RIM BlackBerry, etc.) – even experienced developers are left feeling like beginners [2]. All of the tools, processes, and techniques they have acquired to build, debug, test, and deploy software are suddenly powerless against mobile.

In particular, recently there has been a proliferation of development environments specific to the mobile world. Developers can choose from either *native development tools* for each of the major mobile devices and platforms and *cross-platform environments* (like PhoneGap [5] and Appcelerator Titanium [6]) to create an application that run, at least in principle, across multiple mobile devices and platforms. So new development questions arise for creating simple and reliable applications, including what devices to target and which development frameworks to use.

Based on our recent experience on developing for Blackberry and Android mobile devices, in this position paper we try to provide an answer to two fundamental questions. What kinds of mobile development environments are available, and what are their advantages and disadvantages? Specifically, we compare the two main approaches for developing mobile applications, namely i) by using native APIs and toolkits for the target platform and ii) by adopting cross-platform environments, mostly WEB-based. We present our position on combining preciseness and efficiency of native tools with the flexibility and automation of the cross-platform frameworks to achieve significant boosts in both productivity and quality in mobile application development. Finally, we examine the role of Eclipse in both approaches.

## 2. DEVELOPMENT TOOLS FOR MOBILE APPLICATIONS

Tools vendors have created multiple development environments, but they fall in two main categories: native tools and cross-platform tools.

*Native tools.* They are designed to create applications that run on specific platforms. For example, in the case of Android, this normally means Java and the Android SDK (software development kit). In the case of Apple iOS, development is based on the Objective-C programming language and a toolset for designing and distributing applications include the Xcode IDE for UIs, performance analysis tools, and iOS Simulator. Though based in the past on Java, Blackberry provides for the recent BlackBerry 10 operating system an C/C++ app framework and a plug-in Eclipse. Similarly for other platforms.

*Cross-platform tools.* They provide developers the flexibility to create an application that run across multiple mobile devices according to the ideal principle "write-once-run-everywhere". Examples of cross-platform frameworks that we experienced with are Appcelerator's Titanium and PhoneGap. They are designed to limit the development work and costs to create applications for iOS, Android, BlackBerry, Windows Phone and beyond. An entire sub-industry of development tools and languages exist to develop and deploy a mobile application to multiple platforms.

Basically, most of the existing mobile cross-platform environments are toward open-source world and Web-oriented by incorporating three key technologies: HTML5, Cascading Style Sheets (CSS), and JavaScript. However, they adopt different development approaches. There are web apps that run inside of a browser (either standalone or embedded into a container to more closely mimic a native application). This approach is adopted, for example, by PhoneGap. Phone-Gap uses HTML5 inside of a WebView[1] on the device. It essentially creates a mobile web app that sits inside a native application wrapper. The web code is packaged with a library that bridges web code to native functionality.

There are also approaches that include their own runtime, like Adobe Air [7]. And there are tools, such as Titanium and Corona SDK[8], that allow you to write the code in an abstracted scripting language, and then generate native code at compile time. Titanium, for example, compiles the JavaScript code into a native binary-converting the JavaScript into native classes and object files (whereas PhoneGap simply renders a WebView with the code being interpreted inside). Though, it is close to pure native mapping, there is still an interpreter running in interpreted mode to allow, for example, dynamic code.

As cross-platform tools, there also exist *Mobile Enterprise Application Platforms* (MEAP) (including Antenna Software Inc. and Kony Solutions Inc.) [3] that have more full-fledged development environments, with a wider variety of traditional tools such as graphical user interfaces, version control, and workflow. They tend to have more integration tools and gateways to third-party services (such as Facebook and Twitter), as well as better technical support capabilities. In addition, they focus on the enterprise segment and strive to incorporate stronger security capabilities, taking into account that the applications will be used to access back-end corporate information.

## 2.1 Cross-platform vs. native app development

Both development approaches has advantages and drawbacks. Table 1 summarizes our comparison of the two approaches according to the following criteria (a subset has been taken and revised from [4]).

*UI User Experience.* Native apps provide a more fluid and responsive interface than cross-platform solutions, especially for animations and gestures. This is because when coding with the indented programming language of the platform you have access to the full device APIs. Though cross-platform solutions offer native APIs to use, it always refers to a limited subset of the device-specific features and often you have to wait until they are released in order to use them.

*Performance.* One key advantage of using native development tools is that applications run more smoothly on whichever mobile devices use that operating system. Instead, the cross-compilation process can sometimes be slower than using native tools for an app. This difference can be easily noted during graphical rendering and animations.

*Device-specific features.* In addition, native tools let developers take full advantages of platform functionality. On the contrary, a cross-platform application serves everyone, but has more limited functionality.

---
[1] A browser screen within the native application that then renders the HTML5/CSS/JavaScript page.

For example, with cross-platform apps, high-end graphics and 3D support is often limited. There are some cross-platform solutions (like Titanium) that compile code into the native language, but none of them compile it to completely native.

*Distribution via app-store.* There may be stringent requirements for admission into public app stores. Apple Inc., for instance, requires that developers submit iPhone mobile digital device applications for testing within Apple to facilitate such compatibility.

In this sense, apps developed using native tools are in general more welcome in app store than cross apps because it is assumed that native apps have been developed by programmers with expertises of the target platform.

*Multiple platforms deployment costs.* With native, when developers want to target multiple platforms, they have to adopt application architecture best practices and use of a common data model to optimize the development effort across multiple platforms. Native requires developers with (not so common) skills for the target platforms.

With cross apps, development costs are reduced. This is perhaps the biggest advantage because a cross-compilation schema allows companies or brands to get an app onto other platforms without having to invest in a team or developer specific to that ecosystem.

*Developers support.* Developers we work with often say that they like to work in native code because it is easier to get help. They can go online to forums and quickly get answers since many people started writing native code for a much longer time than cross-platform code (a more recent technique).

*Security.* Cross-platform apps present more security risks than native apps. The reason is that they inherit the same risks of HTML5. For example, to name a few, application source code is freely available on the mobile device, data cached on the device (within the browser) not properly secured and encrypted, URL security vulnerabilities. On the contrary, operating systems like iOS and Android offer built-in security services like data encryption. Attack techniques such as cookie manipulation and SQL injection to gather sensitive data from back end servers and from the mobile device itself are not possible in a (well-built) native app.

*Timely access to new OS innovations.* A cross-platform framework may not support every feature of an operating system or device. Any new added feature of the operating system is not immediately available on the cross-platform framework you are using. You need to wait it is updated to support those new features.

Table 1: Cross-platform vs. native development

|                                       | Native | Cross-platform |
|---------------------------------------|--------|----------------|
| **UI User Experience**                | high   | low            |
| **Performance**                       | high   | low            |
| **Device-specific features**          | high   | low            |
| **Distribution via app-store**        | high   | low            |
| **Multiple platforms deployment costs** | high | low            |
| **Developers support**                | high   | low            |
| **Security**                          | high   | low            |
| **Timely access to new OS innovations** | high | low            |
| **Code reusability**                  | low    | high           |
| **Design challenges**                 | low    | high           |
| **Availability of programming expertise** | low | high          |

*Code reusability.* Cross app code is considered reusable. Rather than having to write a specific action or a sequence of actions for each target platform, a cross developer can just write the code once and then reuse it on other platforms or in other projects. This is not always true. Often some cross-platform frameworks often use their own subsets of JavaScript, which means that if you want to switch to another platform, that code you wrote before is likely not going to be reusable without refactoring or substantial changes.

*Design challenges.* In native development, design is simplified by the support and services provided by the operating system. The operating system can, for example, notify applications about events such as message arrival and power levels. In a cross-platform environment, developers will need to add such features explicitly.

Moreover, with cross-platform frameworks, developers must design how each feature they need has to be implemented on each target platform. For instance, designing an app for the iPhone is different than designing one for Android; the UI and user experience conventions are different, and touch points and menus work in different ways. Personally, we think a good cross-platform application looks at home on whatever platform it is used on. A bad cross-platform tries to look identical everywhere.

*Availability of programming expertise.* It is widely acknowledged that there are more Web developers than native developers. Since most cross-platform frameworks are based on HTML5 CSS3, they are easy for web developers to jump in and use alongside the calls to more native functions. On the other hand, due to the low availability, native developer skills usually cost more.

## 3. ECLIPSE FOR MOBILE APPLICATION DEVELOPMENT

Some tools, native or cross-platform, offer easy access to plugins and modules that can easily plug into other services or tools including Eclipse. For example, both Titanium and PhoneGap are released as Eclipse-based environments. Providing compatibility with common integrated development environments (IDEs) like Eclipse is to be considered a fundamental feature that enable developers to leverage existing tools and expertise.

The real challenge is to find the next generation of development tools and development processes that make mobile application development as productive and manageable as desktop and web development have been for so long. To achieve these productivity goal, there are five aspects [2] requiring for better mobile development tools and for which the Eclipse-based ecosystem may help:

*Building.* Many platforms means many different "build" requirements for writing and compiling an app for iOS, Android, Windows Phone, BlackBerry etc. The building phase requires at the moment the use of different IDEs, SDKs, and operating systems. Also cross-platform apps, which leverage existing Web skills to reach multiple platforms, require lots of complex and messy configurations for each target OS.

New and improved tools to help mobile developers abstract platform differences and manage the building phase in less time are necessary.

*Debugging.* Mobile app debugging is based on mobile operating system emulators that implies software being written on a PC, run on a device, and then debugged from the PC. This is quite sufficient. Tools that makes mobile apps painless to debug on mobile devices are necessary.

*Testing.* Once an app is built, it needs automated tests to ensure it works properly before updates are shipped to app stores and users. For conventional applications this is a relatively straightforward task and a great variety of testing automation tools and techniques exist. This not the same for mobile apps due to the wide variability of today's mobile devices. Mobile app testing needs to happen not only on many different operating systems, but on many different physical devices.

Recently, tools and "cloud device labs" for testing are emerging, but much more is still needed to make it productive to record, playback, and manage tests across devices. For example, when using BlackBerry WebWorks and PhoneGap APIs, a sophisticated emulator called Ripple is avail-able as multi-platform mobile environment emulator that is custom-tailored to mobile HTML5 application development and testing. The Ripple emulator is an extension of the Google Chrome browser that allows you to quickly see how your application looks and functions on multiple mobile devices and platforms in a browser-like environment. You can use the Ripple emulator to perform JavaScript debugging, HTML DOM inspection, automated testing, and multiple device and screen resolution emulation in real-time without redeploying the application or restarting the emulator.

*Deploying.* Deploying mobile apps on app stores requires many manual steps. The problem is even more challenging if the destination is not a public app store but a private group of users. These last must find their own path to employee devices. Tools that help and automate the delivering of multi-platform deployment mobile apps are required.

*Monitoring and Optimizing.* Unlike websites that live on servers or desktop applications that live on relatively stationary PCs, mobile apps get around and once deployed, they are out of your control and destined to be abandoned.

To understand what your app is doing and why, tools that help developers to monitor it in-place to final users are necessary. In particular, developers need to monitor usage and performance, watching for common user problems and those characteristics to fine-tune accurately and productively after initial use. A nice tool toward this direction is, for example, the plug-in for Google Analytics provided by PhoneGap that allow to get reports about the usage and navigation of an app.

## 4. CONCLUDING REMARKS

The application development approach developers have to choose really depends on the application requirements itself. Will it be browser-based, with little or no data saved on the mobile device? Does it require capabilities native to the operating system? Does it require security features or support capabilities?

However, by the comparison we provided in this paper, we can also conclude that both approaches have advantages and disadvantages shortly summarized in Table 1. According to the chosen criteria, we can conclude that native and cross-platform development approaches are complementary. We believe these two approaches can be combined showing how the advantages of one can be exploited to cover or weaken the disadvantages of the other. In order to combine in a tight way preciseness and efficiency of native apps with flexibility and automation of cross-platform apps, conventional software architectural design patterns may be adopted and revised to adopt an hybrid development approach. For example, by using the *layer pattern*, one can provide a reach UI natively and a functionality core by a cross-platform. The decision to use in a layer one or the other relies on how deeply developers want to link the application with the underlying operating system, as capabilities in one operating system may not be available in another.

Moreover, developers have to guarantee all those desired software qualities (reliability, efficiency, maintainability, etc.) also for mobile apps. To this purpose, the integration of mobile development tools with existing or new Eclipse-based tools for building, debugging, testing, deployment and optimizing is fundamental.

# Misura della cooperazione per lo sviluppo su piattaforma Eclipse: lo studio del caso di collaborazione fra le università Federico II e Ohio State University at Stark

Paolo Maresca
Dipartimento di Ingegneria Elettrica e
Tecnologie dell'Informazione (DIETI)
Università di Napoli, "Federico II"
+0397683168
paomares@unina.it

Angela Guercio
Department of Computer Science
Kent State Univ. at Stark, Canton, OH
aguercio@kent.edu

Lidia Stanganelli
Dipartimento di Ingegneria Elettrica e
Tecnologie dell'Informazione (DIETI)
Università di Napoli, "Federico II"
ldistn@gmail.com

## SOMMARIO
Le comunità di pratica costituiscono una portante dell'innovazione di prodotto e di processo, in un contesto di depressione economica esse possono trasformarsi in leva per lo sviluppo di idee per piccole e medie imprese. Esse costituiscono un valore per la nostra società. Questo lavoro ha lo scopo di fornire un contributo teso ad incrementare tale valore attraverso lo studio di un modello di misura, basato su GQM, della cooperazione fra studenti appartenenti a corsi di università diverse (Università di Napoli Federico II e Kent State University at Stark su due corsi di computer science: Programmazione I e Computer science II Data structures and abstractions) che fornisca un piano di misura per verificare l'avvicinamento di comunità di pratica apparentemente distanti fra loro. Il modello promuove il trasferimento di conoscenza fra i membri delle comunità di pratica aperte (OCoP) che è amplificato quando vengono adoperati strumenti di cooperazione innovativa come Eclipse e Jazz.

## Categories and Subject Descriptors
K.3.1 [**Computers and education**]: Computer uses in education – *collaborative learning, distance learning*.

## General Terms
Measurement, Experimentation.

## Keywords
Collaboration, e-learning, learning models, community of practice, cooperation.

## 1. INTRODUZIONE
Generalmente una Community of practice (CoP) ha una sua mission (es. FERRARI), tende a delimitare i propri confini attraverso degli elementi che la caratterizzano (campo di conoscenza, risultati raggiunti, uomini rappresentativi, argomenti di ricerca in cui è forte, etc) scambiando conoscenza quasi esclusivamente con persone della stessa CoP divenendo di fatto una comunità chiusa con una rete di innovazione chiusa (closed innovation framework).

Di rado le comunità si aprono, ma quando lo fanno, hanno una visione aperta dell'innovazione (open innovation network): le idee cominciano a circolare e producono altre idee in un circolo virtuoso di innovazione aperta.

È evidente, posto che si abbia l'interesse a che due o più CoP cooperino, esse abbiano un alto potenziale di scambio di conoscenza ma è molto difficile che siano in grado di attivare questo flusso di conoscenza da sole. Metaforicamente è come l'acqua, che scorrendo, trova la via a minor resistenza per far fluire la conoscenza fra due CoPs. Il problema è che non riusciamo a misurare oggettivamente il grado con il quale queste si avvicinano o si allontanano. Lo possiamo solo verificare a posteriori osservando i risultati della loro cooperazione. Questa incapacità ci costringe spesso a vedere la morte di comunità promettenti senza riuscire a fare nulla o meglio senza avere la possibilità di intervenire per rivitalizzarle e allungarle il tempo di vita.

Da più parti ormai si osserva come le comunità di pratica aperte (Ocop) hanno un modo spontaneo di cooperare attraverso strumenti di social networking. Basta guardare ai grandi passi in avanti effettuati in molteplici settori dello sviluppo del software open source nel quale lo sviluppo è diventato cooperativo ed in settori della formazione universitaria di alto livello dove l'apprendimento è diventato cooperativo riuscendo a rimuovere quegli ostacoli alla cooperazione che nascono dalla insana competizione individuale.

Molti aspetti dei modi con i quali le comunità di pratica si avvicinano, si allontanano, producono, nascono e muoiono non sono ancora stati ben compresi. Riteniamo che molto è dovuto sia alla complessità che alla dinamicità con la quale una comunità si muove. Tale dinamicità richiede che un insieme di individui sia: raggiungibile, osservabile e controllabile. Insomma che la comunità sia misurabile in maniera oggettiva anche attraverso gli asset (manufatti) che gli individui producono e che le sue misure siano espresse attraverso numeri o attributi riconosciuti essere oggettivi. In particolare il lavoro proverà a disegnare attorno ad una Ocop un sistema di misura oggettivo il cui obiettivo sarà quello di poterne capire di più anche circa altri aspetti (vision future, prospettive, sicurezza, etc). Riteniamo che una strada potrebbe essere quella che asseconda la naturale e spontanea consonanza delle comunità che manifestano attraverso il raggiungimento di uno o più obiettivi [12]. Per verificare il raggiungimento degli obiettivi relativi a una comunità di pratica, in questo lavoro, proponiamo di applicare il paradigma del Goal/Question/Metrics [1,2,3].

## 1. Il Goal/Question/Metrics (GQM)

La complessità della misura su una rete di individui appartenenti ad una comunità di pratica aperta è evidente a tutti. Essa è legata alla raccolta di numerosi dati che costituiscono il risultato di una misura in funzione degli obiettivi prefissati. Il problema è quello di ridurre tale raccolta a ciò che è veramente indispensabile perché la raccolta dati è costosa ed è difficile a causa della dinamicità del sistema. Il processo di raccolta dati non è normato da standards mentre invece esiste un approccio sviluppato presso i laboratori di ingegneria del software dell'università del Maryland - USA il cui artefice è Victor Basili. Tale approccio e denominato Goal/Question/Metrics (GQM) [1,4,5].

Il GQM rappresenta un'ottima linea guida per implementare un processo di misura [9,10]. Il paradigma si basa su un approccio sistematico e top down che parte dall'identificazione e definizione degli obiettivi e termina con la raccolta dei dati di misura. Esso è molto pragmatico e sancisce il principio per il quale non esistono programmi di misura standard ma solo linee guida che possono essere usate per sviluppare una metodologia e quindi un programma di misura atto a soddisfare tutte le esigenze.

Il paradigma di misura GQM si articola secondo i seguenti due assi portanti:

- La definizione di un processo di misura di un sistema deve partire dagli obiettivi (goals) della Ocop.

- Il processo di misura è strutturato in tre fasi distinte:

    a. La prima di tipo *concettuale* serve ad identificare gli obiettivi (per esempio usando attività di brainstorming e brain writing durante la loro identificazione) che si vogliono raggiungere.

    b. La seconda di tipo logico attiene alla formulazione di domande (questions) direttamente derivabili dagli obiettivi di cui al punto a.

    c. La terza è di natura *operativa* e consiste nella rilevazione di metriche o dati che servano a verificare il raggiungimento degli obiettivi di cui al punto a.

    d. La quarta è di modalità *operativa* e consiste nella memorizzazione dei valori rilevati o calcolati allo scopo di usarli o riusarli per rimodellare il processo di misura stesso o rielaborare la misura dello stesso obiettivo a distanza di tempo.

Il GQM è un metodo per definire e condurre un programma di misura che rende altresì possibile il miglioramento dello stesso [6,7,8,11]. GQM è stato proposto e applicato come tecnica sistematica per quantificare e sviluppare programmi di misura per processi e prodotti software, tuttavia è stato dimostrato che il paradigma di quantificazione degli obiettivi è applicabile ad ogni tipo di programma di misura (non solo software) [9].

L'idea alla base di tale approccio è che non esista un metodo universale per misure un prodotto o un processo, esistono delle buone pratiche che sono per lo più ascrivibili anch'esse a processi orientati agli obiettivi (goal-oriented) cosicché la collezione dei dati o delle misure non è più un'attività fuori controllo ma imbrigliata in un processo quantificabile e tracciabile attraverso una documentazione esplicitamente prodotta. Questo è il motivo per il quale si ritiene che esso è applicabile a Ocop.

## 2. L'esperimento: La cooperazione accademica

L'esperimento vede la creazione di una CoP virtuale che unisce due istituzioni accademiche che condividono obiettivi comuni: la Facoltà di Ingegneria dell'Università degli Studi di Napoli "Federico II" (UNFII), e il Dipartimento di Scienze della Kent State University a Stark, Ohio, Stati Uniti (KSU). L'esperimento è stato eseguito nel semestre autunnale 2012 e ha coinvolto circa 100 persone di UNFII e circa 20 di KSU. Esso oltre a produrre risultati tangibili, che consistevano in una raccolta di diversi manufatti e lo sviluppo di un grande progetto, è stato l'occasione per la valutazione del processo di formazione erogato. La valutazione è stata effettuata alla fine dell'esperimento, invitando gli studenti a rispondere a un questionario. Le risposte sono state ricondotte a una scala di 5 valori (espressi in termini di rapporto tra aggettivi e numeri) che sono stati utilizzati per misurare i risultati e analizzarli obiettivamente. Fig. 1 rappresenta ogni domanda con il suo valore medio su una stella a 22 rami di un diagramma di Kiviat. Questo diagramma è utile per evidenziare i punti di forza contro le debolezze del nostro progetto.

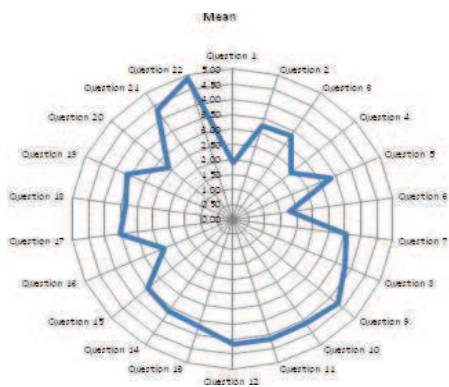

**Figura 1. Diagramma di Kiviat del processo ETC*plus***

Per produrre una valutazione obiettiva abbiamo anche applicato l'approccio GQM ai gruppi di programmazione nel corso del progetto ETC*plus*. Il processo GQM passa attraverso stadi che vanno a definire le misure e i prodotti intermedi generati da questo processo. In particolare abbiamo considerato i seguenti tre obiettivi:

Goal 1: Il miglioramento dell'uso di strumenti di collaborazione per lo sviluppo collaborativo;

Goal 2: Il miglioramento dell'efficienza della collaborazione tra i membri del team per lo sviluppo collaborativo;

Goal 3: Il miglioramento dell'efficienza dello sviluppo nel contesto di sviluppo collaborativo.

La tabella 1 riassume i risultati delle analisi eseguite attraverso GQM nel progetto ETC*plus*. Per semplicità la tabella mostra il valore rilevato, il valore atteso, e la percentuale di corrispondenza conseguita per ciascun obiettivo. La percentuale di corrispondenza indica un valore sufficiente/buono ma non è molto alta per ciascun obiettivo. Il valore più basso si raggiunge con il goal 2. Ciò significa che la collaborazione raggiunta dagli studenti nell'ambito del progetto ETC*plus* è sufficiente, ma deve essere migliorata. Un risultato leggermente migliore si ottiene nel raggiungere il goal 3. Ciò significa che, anche qui, vi è spazio per miglioramenti nello sviluppo di codice, così come nella produzione di codice in termini di completamento del lavoro degli elementi assegnati. Un migliore risultato invece è raggiunto nell'uso degli strumenti di collaborazione (goal 1). In questo caso, la mancanza di esperienza nell'uso di questi strumenti è in parte stata compensata dal buon lavoro dei coordinatori, che hanno fornito agli studenti una grande quantità di materiale di supporto di alta qualità.

**Tabella 1. GQM applicato a ETC*plus***

| Goals | Detected Value | Expected Value | Percentage of correspondence |
|---|---|---|---|
| Goal 1 | 68 | 93 | 73.1 |
| Goal 2 | 52 | 84 | 61.9 |
| Goal 3 | 59 | 87 | 67.8 |

Il goal 2 rappresenta il collante naturale tra il questionario erogato e il GQM. Anche se nel questionario le metriche per la misura della collaborazione non sono state prodotte, abbiamo empiricamente osservato che il sistema (cioè la CoP), lasciato libero di evolversi, ha raggiunto un equilibrio durante la collaborazione e ha prodotto risultati, come qualsiasi sistema vitale [13, 12, 14]. Il GQM invece ha generato un sistema ibrido che mescola una collaborazione guidata con la spontaneità della comunicazione. Sicché abbiamo ritenuto che il GQM fosse non solo il modo per superare la soggettività del questionario ma servisse a migliorare il processo di collaborazione in ETC*plus*. ETC*plus* mostra miglioramenti rispetto a ETC e rileviamo spazio per ulteriori miglioramenti, quando guardiamo i risultati della Tabella 1 relativamente al goal 2. Gli autori ritengono che si debba lavorare sulla consonanza dei mini-sistemi vitali formati dai gruppi per aumentare la collaborazione [12]. Si consideri che al progetto hanno partecipato studenti di lingue e università diverse che sono stati letteralmente "scagliati" in compiti comuni e ci aspettavamo "risonanti" senza precedente coordinamento. Tuttavia la risonanza è il risultato di un processo di collaborazione che si è evoluto. La risonanza è simile a un gruppo di musicisti, individualmente molto bravi, ma che hanno bisogno di prove prima di raggiungere un livello di collaborazione (risonanza) tale da permettere loro di produrre una grande prestazione. La stessa cosa accadrà per gli studenti. Noi crediamo che sia importante che i partecipanti si conoscano l'un l'altro (anche se virtualmente) prima di lavorare insieme.

## 3. RINGRAZIAMENTI

## 4. BIBLIOGRAFIA

# Custom environments for event-driven I/O

Antonio Natali
DISI
Alma Mater Studiorum - University of Bologna
anatali@unibo.it

Sam Golovchenko
CIRI-ICT
Alma Mater Studiorum - University of Bologna
redsop@gmail.com

## 1. EXTENDED ABSTRACT

In the era of the *Internet of things* (`IOT`), devices like sensors, actuators etc. generate and require data that must be aggregated and combined with information from virtual worlds, such as business databases and Web-based applications. The real world becomes usually accessible through small low-cost computers (e.g. *Raspberry Pi*, *Beaglebone black* , etc.) and smart, mobile devices (e.g. *Android*-based devices) able to interact with other computers via the network. `IOT` applications require the design and development of distributed and heterogeneous software systems and calls for new programming concepts and for new scalable, standardised infrastructures. In fact, when an application has to manage large amounts of `I/O`, the only way of keeping the application responsive and allowing it to scale, is using non-blocking `I/O`. Moreover, if asynchronous `I/O` is achieved via multi-threaded applications, efficiency problems arise, caused by unnecessary thread-switches or related to network socket optimization (the `C10k` problem [3]).

The current solution is to design applications according to an *event-driven* programming style, which is based on an *inversion of control*, since the control flow is driven by the events and no more by the user-defined program. *Callbacks* (i.e. object methods or lexical closures that have been registered for events) are used to obtain the results from asynchronous computations. The main design principle is to avoid as much as possible the usage of threads by adopting the *reactor* design pattern, and the main rules are: *i)* don't stop the reactor and *ii)* make the callback code as lightweight as possible.

However, in callback-based applications, numerous isolated code fragments can be manipulating the same data and coordinating the callbacks (e.g. when multiple event handling is required or when event depends on one another) can become a daunting task, a problem today known as the *Callback Hell*. A clear introduction to the complexity that arises from event-driven programming both in sequential and concurrent environments can be found in [4]. The reference application here is a set of loosely coupled entities (called *plans*, so to abstract from programs, processes, etc.) that interact cooperatively without knowing very much about each other through the *publish-subscribe* pattern based on *listeners*. The problem is the plan interference that arises when "independent plan interact in surprising ways, creating new cases that are difficult to identify, prevent or test"[4].

The *Callback Hell* problem has promoted the development of dedicated language abstractions to take care of event handling and state changes. Moreover, the *reactive programming paradigm* (for a survey see [1]) has been recently proposed to tackle issues posed by event-driven applications by providing abstractions to express *programs as reactions to events* and having the language automatically manage the flow of time and data dependencies .

The problem is that most of the available libraries and languages for event-driven or reactive programming are related to *functional* languages and to *JavaScript*, the most dominant language in the Web. Moreover, standardised infrastructures do not fully exist today. As a consequence, software designers must develop event-driven applications using different languages/libraries in different environments. For example they could use *Node.js* (based on the Google's V8 JavaScript Engine) together with *Flapjax* (for reactive abstractions), or *Tornado* (the *Python*'s answer to *Node.js*) or *Scala.React*, an extension of *Scala* that provides reactive programming abstractions in place of the observer pattern. The `Vert.x` system [7] [1] is a polyglot application platform for the `JVM` that provides an asynchronous programming model for writing scalable non-blocking applications whose components can be written in *JavaScript*, *Ruby*, *Groovy*, *Java* or *Python*. For example, a event-driven `HTTP` Server written in `Vert.x` is:

```
─────── HTTP Server in Vert.x (Java version) ───────
public class Server extends Verticle {
 public void start() {
 vertx.createHttpServer().requestHandler(
 new Handler<HttpServerRequest>(){
  public void handle(HttpServerRequest req) {
   String file=req.path().equals("/")?"index.html":req.path();
   req.response().sendFile("webroot/" + file);
 }}).listen(8080); }}
```

Thus, a `Java` component in `Vert.x` reproduces a `Node.js` style of programming, by using objects as callbacks.

In this scenario, our intent is not to reproduce already available programming styles or define a completely new language like [6], but to provide an open and extendible programming environment for the design and implementation (actually in `Java`, but not limited to `Java`) on small computational devices of event-driven, platform-independent applications. We start from the idea that events should be explicitly declared and explicitly managed in order to highlight the application architecture in terms of event-interacting components[2]. To this end, we provide:

1. An infrastructure (written in `Java` and named `RTEM`: *Runtime Task/Event Manager*) that supports a "classi-

---

[1] `Vert.x` moved to *Eclipse Foundation* in August 2013.
[2] Each event ha a message content of type `String` in order to increase the interoperability with non-`Java` applications.

cal" event-driven model by executing a *event-loop* that manages two types of computational objects: *EventHandlers* (similar to callbacks) and *Tasks* (similar to plans). Both *EventHandlers* and *Tasks* can raise events and react to events. To overcome the *Callback* problem, an *EventHandler* is conceived as a terminating computation modelled as a state machine in which actions and state transitions are performed as conventional procedure calls. A *Task* is defined and implemented as a finite state machine in which state transitions returns the control to the `RTEM`.

2. An extension of the basic infrastructure in which the events are stored within the `RTEM` so that they can be logically consumed by tasks even if they are not explicitly registered for the events when they are raised. This is a quite relevant (and perhaps questionable) extension to the basic event-programming model, since it introduces the possibility of *time-uncoupled* interaction. But, from our point of view, this model is just one of the ways of conceiving the communications among the components of distributed and heterogeneous software applications, and could include the event-driven programming model in the context of the general context introduced in [5].

3. A custom language called `ECSL` (*Event Contact Specification Language*), defined and implemented with the *Xtext* [2] technology. This language allows an application designer to specify in a declarative way the events and the application architecture along three main dimensions: structure, interaction and component behavior. Thus, a `ECSL` specification is a model (instance of a custom meta-model) in which the behavior of components is expressed in terms of state-machines.

4. A custom *Eclipse* `IDE` (called `Event-ide`), that automatically generates the software layer that couples the application logic with the underlying `RTEM` platform.

The current version provides basic mechanisms that allow us to avoid the usage of thread and the *Callback Hell* with reference to classical problems., like recursive event-driven computations (e.g. the evaluation of Fibonacci's numbers), event-driven `HTTP/TCP` servers, event-driven components with both a proactive (e.g. counting) and reactive (e.g. react to a `stop` event) behavior (the problem addressed in [6]), or applications that require plan coordination in the sense introduced in [4]. For example, the `ECSL` specification of the `HTTP` server is:

```
           ECSL specification for a HTTP server system
EventSystem  httpServerExternalIO -basic
        /* - Declaration of the components - */
EventHandler connectionHandler ;
Task server -nostart ;
        /* - Declaration of the events - */
Event connection ;
Event startServer ;
Event endRW ;
        /* - Declaration of the external components - */
External ReaderWriter raising endRW ;
External ServersocketExternal raising connection ;
        /* - Declaration of the event-based interactions - */
handlerHandle: connectionHandler handle connection ;
handlerRaise:  connectionHandler raise startServer ;
serverHandle: server handle endRW ;
        /* - Declaration of the behavior of the components - */
BehaviorOf server{ ... }
BehaviorOf connectionHandler{ ... } //creates/starts a server
```

This `ECSL` model can be informally represented in graphical form as follows

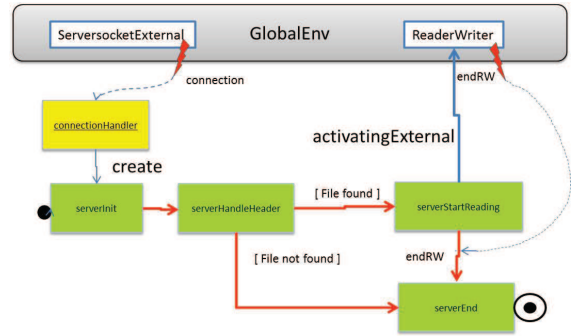

The impact of the custom meta-model on problem analysis and design is the main issue here, once the `RTEM` has proved to be sufficiently efficient (benchmarks are in progress).

A library that implements the `RTEM` and the a couple of Eclipse plugins that implement the `Event-ide` for `ECSL` are our current tangible final products [3]. But thanks to Eclipse and to the model-driven software development methodology supported by `Xtext`, this project is intended to be just the first element of a possible family of custom environments able to express and support more evolute forms of reactive programming mechanisms like component dependencies (so to automatically avoid the *glitch problem* described in [1]) or advanced reactive constructs (e.g. event-stream merging provided by Flapajax) that at the moment must be explicitly handled at application level. The main issue here is the cooperative work between *application designers* and *system designers* (technology experts) in the stream of [5] in order to reuse the same logical (meta)models on different computational platforms, by progressively evolving from single-processor, single-node systems to multi-processor, multi-node environments.

---

[3]See the site `https://137.204.107.21/contact`.

# STS2Java: An Eclipse Plugin for Early Assessment of Service Performance Based on Simulation

Claudio A. Ardagna

Dipartimento di Informatica
Universita degli Studi di Milano
Crema (CR), 26013, Italy
claudio.ardagna@unimi.it

Ernesto Damiani

Dipartimento di Informatica
Universita degli Studi di Milano
Crema (CR), 26013, Italy
ernesto.damiani@unimi.it

Kouessi A.R. Sagbo

Dipartimento di Informatica
Universita degli Studi di Milano
Crema (CR), 26013, Italy
kouessi.sagbo@unimi.it

## ABSTRACT
Since the emergence of the model-driven development paradigm, there has been a significant effort towards the integration of solutions for the assessment of software performance in the early phases of the software development process. Along this line, we have proposed a framework based on simulation that estimates the performance of web services modeled as Symbolic Transition Systems (STSs). Our framework uses the STS-based model of the service under evaluation to automatically produce a simulation script for performance estimation. In this paper, we present *STS2Java*, an implementation of the framework as a plugin for Eclipse IDE, which produces Java-based simulation scripts.

## Keywords
Eclipse plugin, Performance, Simulation, Web services

## 1. INTRODUCTION

The evaluation of software performance is a complex and time-honored problem, which has been exacerbated by the success of the Service-Oriented Architecture (SOA) paradigm and web services [2, 4]. Today, in fact, there is the need of evaluating the performance of highly dynamic and evolving services at design time, by integrating a performance analysis step into the development cycle. To this aim, in [1], we presented a model-based approach that relies on a Symbolic Transition System (STS)-based representation of services to evaluate their performance. In particular, we proposed a framework that extends the standard STS definition [3] with transition probabilities and delay distributions, and uses the extended STS-based models to generate simulation scripts for service performance estimation. In this paper, we present an implementation of our solution as a plugin, called *STS2Java*, within the Eclipse IDE (www.eclipse.org). Our plugin supports developers and expert users in the generation of ready-to-use Java-based simulation scripts, starting from an XML-based encoding of the STS-based model of the services. Simulation scripts are then used for an early assessment of service performance, and to negotiate and evaluate SLAs on service performance.

## 2. OUR FRAMEWORK

Our framework proposed in [1] evaluates the performance of a web service by measuring some performance indicators like service time and response time. Figure 1 shows a simplified version of this framework. Basically, the model of the service is first built from the service interface and code. The model is then given as input to our framework that produces a simulation model by adding transition probabilities and delay (waiting time) distributions to transitions representing internal service operation tasks. The transition probabilities model the normal execution flow of the service. A delay distribution represents the distribution of times needed to complete the task represented by a state transition, and can be adjusted on the basis of coarse-grained information retrieved by testing on the total execution time of service operations. The simulation model is then encoded in XML and used to generate a simulation script that is executed to measure the performance indicators of the service, when the service code is not available. The outputs produced by the execution allow to estimate the behavior of the service.

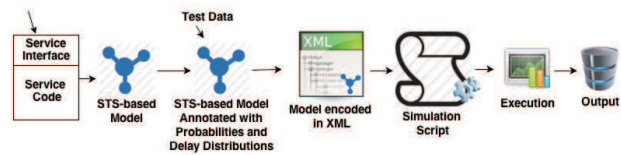

Figure 1: Our simplified framework for simulation script generation

## 3. SIMULATION PLUGIN: *STS2JAVA*

This section presents plugin *STS2Java* for simulation script generation and some experimental results of its execution.

### 3.1 Overview

The Eclipse plugin *STS2Java* (available at http://sesar.dti.unimi.it/sts2java/) helps developers and expert users in assessing the behavior of a service at design time, via simulation. It is compatible with Eclipse 3.4 and JavaSE-1.6, and implements two main components: *i)* a *parser* that checks the validity of the XML encoding of the STS-based simulation model; *ii)* a *generator* that generates a Java-based simulation script. The execution of the generated script provides an estimation of the performance of the service under evaluation. This performance can be compared with results retrieved by testing the real service, to evaluate the quality of our simulation approach.

### 3.2 Execution flow

After *STS2Java* has been installed following a traditional Eclipse installation procedure, entry "STS2Java" is added to

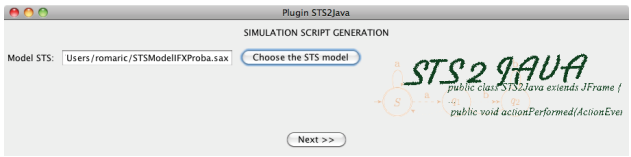

Figure 2: Interface for STS-based simulation model selection

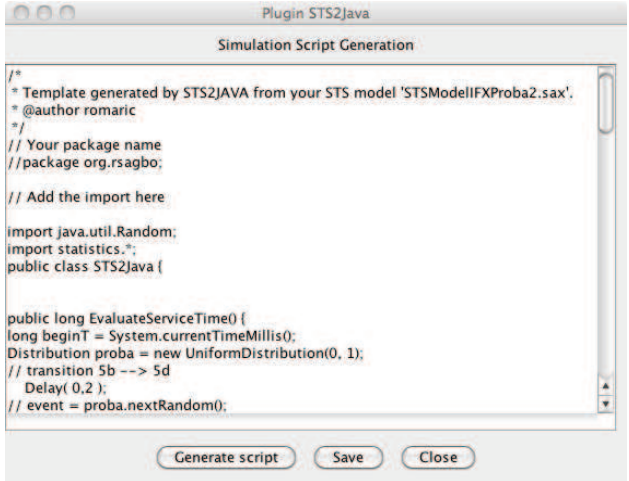

Figure 3: Interface for simulation script generation

```
 1  public long EvaluateServiceTime() {
 2    long beginT = System.currentTimeMillis();
 3    Distribution event = new GenerateRandomEvent();
 4
 5    // transition (5,5a)
 6    Delay(Uniform(0,4));
 7    // transition (5a,5b)
 8    Delay(Uniform(1,4));
 9    Double pevent = event.nextRandom();
10    switch (pevent) {
11    // transition (5b,5c) and (5c,5)
12      case pevent <= 0.1:
13        Delay(Uniform(1,1));
14        Delay(Uniform(1,1));
15    // transition (5b,5d) and (5d,5)
16      case pevent > 0.1:
17        Delay(Uniform(4,7));
18        Delay(Uniform(2,9));
19    }
20
21    return System.currentTimeMillis() - beginT;
22  }
```

Figure 4: Java-based simulation script

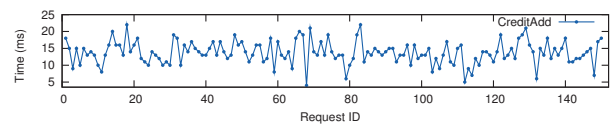

Figure 5: An example of simulated execution times

the Eclipse main menu. Upon starting *STS2Java* by clicking on the new entry in the menu, the interface in Figure 2 is shown to the user. The user can then choose the STS-based model for simulation representing the service to be evaluated, and click on button "Next" to reach the interface for simulation script generation (Figure 3). At this point, the user clicks on button "Generate script" to generate the simulation script and on button "Save" to save it for execution. Figure 3 shows a fragment of a simulation script (i.e., Method **EvaluateServiceTime**) that has been generated by our plugin. We note that the generated script is integrated within Java class **STS2Java** and can be executed within Eclipse IDE to show the service performance and its trend of execution times.

### 3.3 Experimental results

We show the outputs of our plugin by using the reference scenario described in [1], which considers a financial service implementing a reverse ATM. In particular, using *STS2Java*, we generate a simulation script to evaluate the performance of operation *CreditAdd* that allows authenticated users to deposit funds on their bank account.

Figure 4 shows the Java-based simulation script generated by our plugin on the basis of the simulation model in [1]. Method **EvaluateServiceTime** first starts a clock (line 2) and calls, for each activity in the execution flow of operation *CreditAdd*, an instruction *Delay*. This instruction *i)* randomly selects a waiting time (ms) using the probability distribution given as input, and *ii)* implements a Java thread that sleeps for the generated waiting time. We note that a **switch case** statement is generated to model possible alternatives in the execution flow (lines 10–19). Instructions in lines 3 and 9 permit to generate a random number that is used to select a given alternative (lines 12 and 16) following the transition probabilities defined in the simulation model. At the end of the simulation, the clock is stopped and the execution time computed (line 21).

Figure 5 shows an example of performance evaluation given by the execution of the simulation script.

## 4. CONCLUSIONS

We presented *STS2Java*, an Eclipse plugin that allows developers and expert users to simulate the behavior of a service from its STS-based model. Using this plugin, the performance analysis step can be integrated into the development cycle of software/services and allows an early assessment of their performance. Our future work will provide an integrated environment, supporting the automatic execution of the simulation scripts, with an interface showing the trend of the execution times and performance for the simulated services.

# "Il tesoro di Crema": a QR Code Web Treasure Hunt powered by Eclipse

Mario Bolignano
Università degli Studi di Milano
Computer Science Department
Crema, Italy
mario@bolignano.it

## ABSTRACT

The application described in this work is the first public implementation of our framework that aims to create, organize and manage web-based "Treasure Hunt" games using QR Codes. The use of an IDE as Eclipse for Java EE Developers has played a strategically role firstly to simplify the setup and development of a framework for web applications and then to instantiate from it the specific case of study.

The proposed application takes part to the "Una App per Crema" contest organized by "Associazione Cremasca Studi Universitari". The case of study demo is available at http://crema.mbns.it

## Keywords
Web Application, Framework, QR code, Treasure Hunt, Tourism.

## 1. INTRODUCTION

Nowadays, mobile communication makes use of phones that vary in features and specifications: while low-end ones that just support basic telephony still exist, most of mobile handsets currently sold are so called "smartphones" [2], devices with an advanced operating system and hardware, supporting a wide range of software applications. As a result of technological improvement in cellular networks as well, internet connection is also made available to mobile phones, and smartphones are naturally the main recipients of this technological advantage.

One of the more common features made available on almost every smartphone is the geographical localization of the phone itself and therefore of its owner. This is made possible, alternatively, thanks to a GPS antenna embedded in the handset if existing, or other means, such as GSM network cell localization, detectable WiFi access points available in the area and so on.

Based on this feature, a lot of applications make use of the detected user coordinates to "check in" her/him at nearby locations (often commercial venues or touristic attractions) in order to allow her/him to meet friends, exchange reviews, collect some kind of credits and other purposes.

This schema, anyway, doesn't reflect the fact that just being "near" a place doesn't mean having really "discovered" it, and since the developed framework aims to create and to properly manage a Treasure Hunt, another mean of "check-in" is required to fully guarantee that the goal has been reached, especially if the place is intended to be hidden or disguised.

Making use of another feature that is almost essential for a smartphone, i.e. the camera, the idea underlying the proposed framework is to make use of QR codes: they would be placed nearby the intended locations and carry the information that allow the user to properly check in that specific place, if and when some preset conditions occur (for example: user enrollment to the hunt, user proximity to the location, proper order in the path…).

The framework has been designed to support a generic concept of treasure hunt as a game, but the discovery process can also drive a website created to support local tourism or business.

The presented case of study based on this framework has been implemented as a technological aid to tourism for the city of Crema, Italy. In the following sections we will describe the case of study architecture and the essential role of Eclipse in tailoring the final application.

## 2. CASE OF STUDY

The proposed application aims to promote and enrich the tourists' experience in the city of Crema, providing visitors with an innovative and captivating way to discover the territory, i.e. a "treasure hunt".

The innovative aspect is guaranteed by the use of the QR Code technology, commonly known as "matrix barcode" or "two-dimensional barcode". A barcode is an optically machine-readable label that is attached to an item and that records information related to that item. A QR code consists of black modules (square dots) arranged in a square grid on a white background intended to be translated and interpreted through, for example, a smartphone camera [1].

The underlying idea is that the different Points Of Interest (POIs) in the city of Crema can be organized in several thematic paths, i.e. cultural, touristic, gastronomic etc., so that the POIs can be viewed as individual stages of a specific treasure hunt game. A QR Code will be physically located nearby the referring POI. By scanning the code with a smartphone or a tablet connected to the internet, the user will access the web application that shows information about the current POI.

Unlike a traditional treasure hunt game, in order for the user to find the next stage, the application does not provide a clue to solve: it could be boring or impossible to solve for a tourist; the directions to reach the next stage (POI) in the specific thematic path are showed instead. Moreover, the directions can be designed and presented so to induce users to follow the most befitting route, not just the shorter one, in order to enrich the experience of the territory discovery.

The application tracks the user movements so that he/she can display the already reached POIs on a map; this kind of visualization helps them to better recognize the reached places and the followed route and eventually plan future trips.

The application also provides the management tools to create and organize the treasure hunt games via an "admin" panel. The game promoters can define different POIs, providing images and information; they can create and organize distinct thematic paths, associate POIs to them, and finally generate the QR Codes to print and physically place nearby the POIs.

A working prototype of the application is currently available at http://crema.mbns.it. Access to the site is subject to enrollment via the specific form; an Italian mobile phone number enabled to the SMS service is required. Some demonstrative thematic paths and their QR codes are available at: http://mbns.it/democrema.

## 3. WEB ARCHITECTURE

The application architecture relies on a three-tier architecture, composed by a data storage at the back end, a business logic layer, and a visualization layer at the front end, respecting the Model-View-Controller design pattern. To achieve modularization, communication among or intra modules is carried out through standard open protocols for exchanging data formats like XML, JDBC etc.

### 3.1 Data Storage Layer/Model

This component is responsible for storing the application data, making them persistent and retrieving them on request. It relies on the PostgreSQL DBMS. This layer communicates with the upper one by means of JDBC protocol, feeding the Java Beans classes representing the database entities in the Business Logic layer.

### 3.2 Business Logic Layer/Controller

This component is composed by a set of Java Servlets that carry out user requests generated in the view component:

- *LoginServlet*: it manages user registration and sessions;
- *GameServlet*: it manages the treasure hunt game actions, checks the validity of the parsed QR code and manages the user statistics.
- *AdminServlet*: it manages the creation and editing of POIs, thematic paths and QR Code generation by an admin user.

### 3.3 Visualization Layer/View

The main goal of this layer is to present the game information to the users and to notify the underlying layers about queries and requests issued by them.

The layer comprises a set of dynamically generated JSP pages. The main pages are:

- *Scopri.jsp*: this page displays information about the scanned QR Code POI. It also provides, to authenticated users only, information about the path directions to the next POI.
- *Esplora.jsp*: this page welcomes an authenticated user not yet registered to a thematic path, listing the available ones. Access to a path happens by finding any QR code belonging to it, since paths are designed to be circular, with no preset starting point.
- *Admin.jsp*: this page allows creation of POIs and thematic paths and management of existing ones.

If at any moment the requested page is not consistent with the user status (e.g. enrollment status, active path, query parameters) a redirection mechanism takes care of providing the more suitable page for the current status, showing error messages when needed.

## 4. ECLIPSE IDE

The use of an IDE as Eclipse, and its extension for Java EE web developers, played a strategically role in the development both of the game framework and of the specific web application for Crema treasure hunt. In the following sections we will describe the specific features that turned out to be essential for a faster and simpler engineering of the presented work.

### 4.1 Web Environment

Building a web application means dealing with a set of technologies and tools that must be set up for their cooperation through scripts, xml configuration files etc. These are error prone operations for a developer and the use of an IDE automatizing such operations reduces errors and programming time.

The Eclipse Java EE extension for web developers allows creating a *Dynamic Web Project* that takes care of the configuration aspect between the various required tool (e.g. a servlet server like Apache Tomcat) for example by automating the registration of a servlet upon creation into the web.xml configuration file or offering the stubs for the more commons web related methods, or making it possible with a few mouse clicks to export the whole project as a single file to be transferred and deployed on the production server.

### 4.2 User-friendly Code Editor

The described architecture is easily organized and managed thanks to Eclipse natural support for the MVC pattern, which allows developer to easily create servlets, beans and JSP/HTML pages, to edit or replace them individually while still retaining an overall solidity by means of powerful features of code refactoring.

In fact, the Eclipse built-in code editor significantly improves the coding experience not only offering basic features like syntax highlighting for several languages (mainly useful for a web project which concurrently includes Java, JSP, HTML, CSS and Javascript code) or automatic code indentation, but including more advanced functions for code refactoring (e.g. auto updating references) which in our case greatly simplified the creation of the case of study application from the treasure hunt game generic framework, specifically allowing a clean transition from a naming convention to another and simplifying the code recycling for the newly added or improved code sections.

### 4.3 Built-in Debugger

"Everyone knows that debugging is twice as hard as writing a program in the first place. So if you're as clever as you can be when you write it, how will you ever debug it?" [3].

Debugging clearly covers a remarkable role in the software development timeline, and the Eclipse web tools greatly empower developers by offering them the opportunity to even debug the web queries that occur during the application execution, allowing the step by step analysis not only of the code itself but of the web environment overall, accurately reproducing the final product behavior so to significantly shorten the time needed for the debug.

## 5. CONCLUSION

In this work a framework to create treasure hunt web games based on QR Codes has been introduced and a specific case of study with touristic purposes has been detailed.

The dual innovating features of the web application lie not only on the use of the ever more widespread QR Codes but also in the unusual perspective of visiting a city as if it was a treasure to discover, without knowing in advance the places and the paths that will be met during the trip.

Eclipse Java EE extension for web developers proved to be a simple yet powerful environment to easily design, develop and maintain both the general framework and the case of study app.

# Knowledge Hound: Progetto di un social network orientato allo scambio della conoscenza

Antonio Esposito
Università degli studi di Napoli Federico II
Via Alcide.De Gasperi, 151
80021 Afragola NA
+39 3343372919
antonio.esposito85@studenti.unina.it

## ABSTRACT
In questo documento si intende affrontare il problema della condivisione della conoscenza e della cooperazione tra gli utenti mediante strumenti telematici e porre l'attenzione sull'ostacolo rappresentato dalla frammentazione delle informazioni e dalla dispersione degli utenti su un numero eccessivo di piattaforme indipendenti.

Si intende infine proporre il progetto di una piattaforma in grado di soddisfare al suo interno tutte le esigenze degli utenti che desiderano usufruire di un ambiente integrato per la collaborazione e la condivisione della conoscenza, allo scopo di illustrare quali sono le funzionalità che un ambiente di questo tipo dovrebbe offrire e quali sono le altre caratteristiche che dovrebbe possedere.

## Categories and Subject Descriptors
K.3.1 [**Computers and education**]: Computer uses in education – *collaborative learning, distance learning*.

## Keywords
Collaboration, e-learning, learning models, community of practice, cooperation

## 1. INTRODUZIONE
La costante evoluzione della tecnologia ed in particolare dei mezzi di telecomunicazione ha drasticamente modificato il nostro approccio alla conoscenza. In ambito accademico ed aziendale si riscontra la crescente tendenza ad affidarsi a motori di ricerca, social network e più in generale a strumenti online per accedere quasi istantaneamente a vaste quantità di informazioni nonché allo scopo di produrre conoscenza in maniera collaborativa.

### 1.1 Problematica
Il principale ostacolo alla diffusione della conoscenza attraverso questi metodi, tuttavia, risiede proprio nell'estrema dispersione, e spesso nell'esclusività, delle piattaforme utilizzate: seguendo le rotte implicitamente definite dagli strumenti indipendenti messi a disposizione da atenei ed aziende, vengono di fatto a crearsi comunità isolate. In una situazione del genere la conoscenza tende a formare dei punti di accumulazione ed il suo trasporto da un punto all'altro non può avvenire senza fatica. In quest'ottica assume particolare interesse lo sviluppo di una piattaforma specific-purpose che riunisca in un unico ambiente tutte le funzioni necessarie alla gestione della conoscenza.

### 1.2 Obiettivi
Tramite il confronto con strumenti già esistenti ed attraverso l'osservazione di comunità attive, si è cercato di individuare quali sono le principali esigenze degli utenti. Si è rilevato che ciò che essi desiderano è disporre di un ambiente in cui possano agevolmente condividere conoscenze e documenti d'interesse comune, mettere in mostra il proprio know-how, reperire informazioni e stringere contatti di collaborazione con altri utenti, tutto ciò in tempo reale e nella maniera più semplice possibile.

Tenendo presenti questi requisiti, si è scelto di sviluppare la piattaforma in oggetto nella forma di una applicazione per cellulari che realizza nella pratica un social network completamente orientato alla gestione ed alla condivisione della conoscenza, ovvero dotato di tutte e sole le funzionalità specificamente necessarie per soddisfare le esigenze osservate e privo di ulteriori features general-purpose che avrebbero il solo effetto di complicare e vincolare le attività degli utenti.

## 2. CARATTERISTICHE
Tutte le funzionalità della piattaforma progettata ruotano attorno ad una manciata di entità chiave che, opportunamente gestite e messe in collegamento, perseguono gli scopi precedentemente illustrati.

### 2.1 Comunità
Si tratta di gruppi, generalmente aperti e comunque sempre visibili, in cui utenti appartenenti alla stessa istituzione, coinvolti in uno stesso progetto o semplicemente interessati allo stesso argomento possono aggregarsi, condividere conoscenza sotto forma di *Oggetti* e dichiarare di aver acquisito competenze mediante un sistema di *Skill*. Sono dotate di sistemi interni e personalizzabili di credenziali che, oltre a suddividere le responsabilità di amministrazione, hanno anche lo scopo di qualificare il ruolo e le attività dei suoi utenti.

### 2.2 Skill
Ciascuna *Comunità* crea un proprio sistema di *Skill*, legate ai suoi campi di interesse, che gli utenti possono acquisire dichiarando all'interno della piattaforma il possesso di uno specifico bagaglio di conoscenze e competenze.

## 2.3 Oggetti

Gli utenti possono inserire all'interno di una *Comunità* uno o più *Oggetti*, (materiale didattico, modelli, esercitazioni, file di progetto e così via) che possono essere scaricati, valutati e commentati dagli altri utenti.

## 2.4 News

Allo scopo di tenere costantemente aggiornati gli utenti sull'evoluzione delle *Comunità* di cui fa parte, ogni evento che avviene all'interno di una *Comunità* viene associato ad uno specifico tipo di *News*, ovvero una notifica che descrive l'evento. Gli utenti iscritti alle *Comunità* hanno la facoltà di scegliere a quali eventi sono interessati e quindi di ricevere in tempo reale una notifica della loro occorrenza.

## 3. CONCLUSIONI

Il progetto presentato non ha certo la pretesa di costituire un nuovo standard per la condivisione della conoscenza mediante strumenti telematici; intende semplicemente dimostrare l'utilità di poter disporre di un'unica piattaforma che abbia la capacità di riunire in sé differenti realtà interessate alla gestione della conoscenza e che una piattaforma di questo genere, per poter risultare efficace, ha la necessità di essere quanto più semplice ed aperta possibile, facilmente accessibile, e di non essere inquinata con funzionalità aggiuntive che si discostano, anche minimamente, dal suo scopo principale.

## 4. ACKNOWLEDGEMENTS

# Utilizzo di GWT per la Generazione di un Editor di Metadati di Mappe di Suscettività

Mauro Fustinoni
Università degli Studi di Bergamo
Facoltà di Ingegneria
mauro.fustinoni@gmail.com

## ABSTRACT
Lo scopo di questo articolo è quello di fornire una presentazione generale delle caratteristiche di GWT e del corrispondente plugin realizzato per Eclipse. Vengono inoltre descritti i possibili vantaggi derivanti dall'utilizzo di tale strumento, riportando, a titolo di esempio, le caratteristiche dell'applicazione realizzata per la piattaforma INSPIRE.

Viene infine presentato brevemente il progetto SISTEMATI con riferimento all'utilizzo di tale strumento, nella costruzione di un'applicazione web basata sulla piattaforma INSPIRE.

## General Terms
Documentation, Design, Theory

## Keywords
Applicazioni web, georeferenziazione, GWT

## 1. INTRODUZIONE

Google Web Toolkit (GWT) è un toolkit di sviluppo per la costruzione e l'ottimizzazione di applicazioni internet complesse [1]. GWT è stato sviluppato da Google è viene attualmente utilizzato da parecchi prodotti e servizi offerti dall'azienda stessa, tra cui Google AdWorks e Google Blogger, per citarne alcuni. Il toolkit è distribuito open-source, completamente gratuito, ed è ampiamente utilizzato nell'ambito della programmazione di applicazioni web per le sue particolari caratteristiche, che lo rendono uno strumento semplice e molto efficace.

## 2. GOOGLE WEB TOOLKIT
### 2.1 Le Principali Caratteristiche

La peculiarità che rende GWT uno strumento di grande successo è il fatto che permette di costruire applicazioni AJAX, anche molto complesse, in linguaggio Java e, successivamente, compilare il codice sorgente in JavaScript altamente ottimizzato per essere compatibile con sostanzialmente la totalità dei browser web diffusi sul mercato. Quindi, non solo permette di programmare facilmente le funzionalità richieste dall'applicazione in un linguaggio tra i più diffusi e usati dalla comunità degli sviluppatori, ma fornisce anche una certa sicurezza sulla successiva portabilità dell'applicazione stessa tra i vari browser esistenti (compresi browser forniti su dispositivi mobile come iPhone e Android), permettendo quindi agli sviluppatori di concentrare i propri sforzi sull'implementazione delle funzionalità desiderate, piuttosto che su questioni legate alla compatibilità degli script generati.

L'SDK di GWT mette a disposizione del programmatore un set di Java API e widget, che forniscono una buona base di partenza per l'implementazione di gran parte delle funzionalità che si potrebbero voler inserire nella propria applicazione. Tuttavia, non si è limitati all'utilizzo degli stessi; infatti, GWT è in grado di accettare e supportare qualsiasi funzione che può essere normalmente realizzata con JavaScript o DOM. Inoltre, fornisce la possibilità di gestire anche codice JavaScript scritto "a mano", quindi non generato dal compilatore, includendolo nella distribuzione finale dell'applicazione in modo integrato con le altre funzioni derivanti dai sorgenti in Java.

GWT permette anche di suddividere il codice sorgente in due parti, rispettivamente per il client e per il server, di modo da poter realizzare architetture software più complete. Le due parti sono scritte entrambe in Java, ma al momento della compilazione, solo la parte di sorgenti riferita al client viene compilata in JavaScript, per permetterne l'esecuzione nell'ambiente del browser. Il codice sorgente riferito al server viene invece compilato normalmente ed eseguito dalla JVM direttamente sul server. Di fatto, GWT consente di realizzare le due parti con un unico linguaggio eliminando la difficoltà di dover lavorare con linguaggi diversi.

### 2.2 GWT Plugin per Eclipse

Insieme all'SDK per lo sviluppo, Google distribuisce liberamente anche un plugin per Eclipse, che viene costantemente aggiornato per essere supportato anche dalle più recenti versioni dell'IDE.

Oltre al compilatore GWT, il plugin integra con l'IDE un ambiente di sviluppo che permette di eseguire e testare direttamente il codice scritto senza la necessità di doverlo compilare immediatamente e senza quindi dover montare l'applicazione compilata su un server web. In sostanza permette quindi di avviare il software su un server virtuale e di visualizzare il front-end dell'applicazione direttamente sul proprio browser, dando la possibilità di effettuare il debug come se si trattasse di un normale applicazione desktop. Questa caratteristica è di indubbia utilità in fase di implementazione, permettendo allo sviluppatore di verificare in pochi passi gli effetti delle proprie modifiche al codice sorgente.

### 2.3 Ottimizzazione del Codice

GWT fornisce anche alcuni tool per l'ottimizzazione del codice generato, in fase di compilazione, che permettono, per esempio, di eliminare parti di codice non utilizzato o comunque, in generale, di snellire il codice sorgente, col fine di ottenere versione finale dell'applicazione più leggera ed efficiente.

Inoltre il compilatore è in grado di dividere in più segmenti JavaScript l'applicazione compilata, separando "moduli" della stessa che sono atti a funzionalità indipendenti e che possono quindi essere trattati separatamente. Questo permette di caricare l'applicazione nel browser in modo incrementale, migliorando notevolmente le prestazioni di start-up per applicazioni di grosse dimensioni.

## 3. LA PIATTAFORMA INSPIRE

Un valido esempio di applicazione web realizzata con l'utilizzo di GWT è costituito dalla piattaforma INSPIRE per la generazione strutture di metadati, secondo lo standard approvato dalla commissione europea. L'applicazione è operativa e raggiungibile all'url http://inspire-geoportal.ec.europa.eu/editor/.

Sostanzialmente, l'applicazione fornisce all'utente un'interfaccia web composta da diversi moduli che permettono l'inserimento di metadati e la successiva costruzione di un file XML per salvarli e renderli utilizzabili per molteplici applicazioni pratiche. I valori inseriti nei campi della GUI producono l'aggiornamento real-time, in modo asincrono, di una struttura XML appositamente formulata e volta a contenere tutte le informazioni fornite dai vari form. L'applicazione permette quindi la generazione ed il download del file XML associato alla struttura, che risulta quindi direttamente a disposizione dell'utente.

Queste funzioni sono eseguite da comandi Javascript che sono stati appunto generati da codice sorgente scritto in Java, tramite il compilatore di GWT. Il risultato è un'applicazione piuttosto fluida, nonostante le dimensioni non siano molto ridotte.

## 4. IL PROGETTO SISTEMATI

SISTEMATI è un progetto, in fase di sviluppo, che si propone di raccogliere e sfruttare informazioni riguardanti rischi di calamità (come gli incendi) e dati storici delle stesse, per poter ottenere dei miglioramenti, in termini di efficacia degli interventi, in caso di emergenza.

### 4.1 Necessità di Georeferenziazione

Una parte di questo progetto consiste quindi nell'avere la possibilità di tradurre enormi quantità di dati, associati alle mappe di rischio per le varie possibili calamità, in dati più sintetici che possano quindi essere più facilmente manipolati, consultati ed interrogati, dando accesso ad importanti informazioni per la previsione dell'evoluzione di particolari eventi che comportano rischi ambientali e alle persone [2].

Ne consegue, tuttavia, che a questi dati "riassuntivi" deve rimanere associata un'informazione relativa alla posizione geografica e al momento temporale a cui i dati stessi fanno riferimento; ossia, è necessario georeferenziarli, integrandoli con metadati che aiutino a catalogarli e organizzarli.

Il codice sorgente della piattaforma INSPIRE, preceden-temente descritta, è open-source e liberamente distribuito. L'applicazione è strutturata in maniera tale da permetterne l'estensione; ossia, è progettata per essere utilizzata come base di partenza per la costruzione di una nuova applicazione che possa permettere l'aggiunta di dati arbitrari alla struttura base XML con cui sono organizzati e salvati i metadati.

La scelta di produrre un'evoluzione di tale software è quindi favorita dalla presenza di funzioni per la gestione e la formulazione di strutture XML, già adattate allo standard vigente, per l'associazione di metadati ai dati arbitrari, come quelli relativi alle mappe di suscettibilità ai vari scenari di rischio.

### 4.2 Estensione delle Piattaforma INSPIRE

L'utilizzo di GWT per la costruzione della piattaforma INSPIRE ha sicuramente facilitato l'estensione della stessa, permettendo di potersi concentrare sull'implementazione delle nuove caratteristiche, sfruttando il riuso di widget e funzioni già implementati per l'applicazione originale; inoltre, GWT ha permesso di programmare direttamente in Java per la realizzazione delle nuove funzionalità sia lato client che lato server, eliminando la difficoltà di dover lavorare con due linguaggi diversi.

Per quanto riguarda, nello specifico, l'interfaccia grafica, è stata necessaria la creazione di una nuova tab, che potesse contenere tutti i form relativi ai dati sui rischi di calamità; tuttavia, grazie all'utilizzo di GWT, non si è dovuto creare "da zero" ogni elemento della schermata aggiuntiva, ma si è potuto riutilizzare parte dei widget già costruiti per l'applicazione base. La gestione dei nuovi elementi inseriti è stata quindi demandata alla infrastruttura software già esistente ed, eventualmente, modificata dove se ne è presentata la necessità.

La piattaforma INSPIRE, pur avendo un buon livello di complessità, non offre comunque un'interfaccia molto dinamica che permetta, per esempio, l'auto-completamento di alcuni campi. Si è quindi dovuto implementare ex novo ogni funzione che richiedesse un aggiornamento più dinamico dei form dell'interfaccia o che permettesse una più semplice customizzazione delle liste di parametri selezionabili all'interno dei vari moduli. L'aggiunta di queste funzionalità è permessa dall'utilizzo delle API di GWT, che forniscono supporto per l'implementazione dei normali comandi eseguibili in JavaScript.

Per accogliere i dati relativi ai parametri delle mappe di rischio, si è successivamente esteso lo schema base XML dei metadati, aggiungendo campi volti a contenere i nuovi valori inseriti. Per sfruttare a pieno la funzione, già fornita da INSPIRE, che consente l'aggiornamento automatico della struttura XML, si è poi reso necessario esplicitare l'associazione di ogni nuovo campo della struttura dei metadati ad un campo dell'interfaccia estesa, eventualmente nascondendo visivamente alcuni di essi, ove si volesse renderli non editabili.

L'inserimento automatico, nella struttura dati XML, dei dati relativi agli scenari di calamità, estrapolandoli direttamente dalle mappe di rischio, ha richiesto invece alcune modifiche più complesse. Si è dovuto infatti demandare questa funzione ad una servlet, risiedente per l'appunto sul server, che, in risposta all'upload dei dati relativi ad un mappa, restituisce i dati trattati e formattati per poter essere integrati con i metadati provenienti dai vari moduli dell'applicazione. Innescando inoltre un nuovo aggiornamento dinamico dell'interfaccia.

## 5. CONCLUSIONI

GWT è sicuramente uno strumento molto utile e relativamente di facile utilizzo nell'ambito della produzione di applicazioni web. Si è inoltre potuto constatare che la piattaforma INSPIRE, basata su questo toolkit, ha fornito una buona base di partenza per l'implementazione delle funzionalità richieste nell'ambito del progetto SISTEMATI.

## 6. RINGRAZIAMENTI

# Plaze

Gianluca Vaccaro
Università degli Studi di Napoli Federico II
gvaccaro90@hotmail.it

**Plaze** è un social network ma con qualcosa in più!

Il mondo dei social network sino ad ora si è limitato a renderci tutti parte di una realtà meramente virtuale.

In questi anni i social network hanno avuto un boom di crescita molto elevato, il loro scopo è quello di avvicinare le persone, il paradosso però è che ci sono persone con migliaia di amici virtuali sparsi per il mondo ma senza amici nella vita reale, che passano le giornate incatenati davanti al proprio pc/smartphone nell'illusione di vivere.

Plaze da un lato accomuna tutte le caratteristiche dei social network, dall'altro li rende più social, evitando l'allontanamento dal mondo reale. L'utente in qualsiasi momento potrà "connettersi" con le persone vicine a lui (ragazze comprese!) vincendo così la timidezza dell'approccio diretto e, dopo aver scambiato due parole virtuali, continuare il discorso fisicamente grazie alla vicinanza. Inoltre, una volta stretta l'amicizia su plaze, l'utente sarà in grado in ogni momento di conoscere la posizione dei suoi amici, quando questi sono on line.

Partendo da una piazza, luogo di aggregazione sociale per eccellenza, i nostri utenti potranno visualizzare le persone che ivi si trovano e a quel punto chattare insieme, condividere attraverso i loro profili le foto ed i ricordi di quel posto, venirsi incontro e stringere amicizie virtualmente ma soprattutto anche fisicamente.

Attraverso l'applicazione, qualsiasi luogo della città potrà avere un suo profilo! Arrivati, per esempio in una piazza, ma anche in una strada, potremo visualizzare le persone e le foto di quel posto e inoltre informarci sulle offerte commerciali e di divertimento che i commercianti potranno inserire sull'app.

La ricerca, oltre al livello locale, può estendersi anche al livello mondiale: potendo l'utente visualizzare tutte le persone che si trovano in un determinato luogo del mondo (entro un raggio di 500 mt. dallo stesso), inserendo il nome della città e della via.

In questo modo l'utente partendo da un semplice indirizzo può farsi un'idea concreta del posto ricercato conoscendo:

- la tipologia di persone che lo frequentato (tramite foto e commenti);
- l'utente può capire se c'è vita notturna;
- se il posto è frequentato o affollato;
- se è caro o economico e altro ancora, per decidere se andarci;
- rintracciare facilmente una persona (e scoprire cosa fa e con chi sta);

In tal modo l'utente non si sentirà mai solo potendo, anche durante l'attesa dei mezzi pubblici, chattare con chiunque gli sta intorno e stringere nuove amicizie non più solo virtuali.

Gli amici di Plaze si suddividono in due schermate:

- una che mostra all'utente gli amici che si trovano nelle sue vicinanze
- un'altra che invece comprende tutti i suoi amici nel mondo con l'indicazione del luogo nel quale si trovano in quel momento (se sono on line, o lo sono stati nei dieci minuti precedenti).

Inoltre l'utente di Plaze può anche vedere tutte le foto condivise dagli altri sia intorno a lui, sia nei posti da lui ricercati e dare un giudizio positivo o negativo su di esse.

Le foto col maggior numero di voti positivi saranno inserite nei "Best", suddivisi in sezioni con le migliori foto della città (visibili componendo il nome della città scelta) e le migliori foto del mondo.

L'utente di Plaze ha poi la possibilità di impostare due diversi tipi di privacy:

1. permettere la visualizzazione delle sue foto solo ai suoi amici oppure a tutti gli utenti.
2. decidere se accettare messaggi da tutti gli utenti oppure solo dai suoi amici.

Inoltre Plaze è il primo social network che si preoccupa anche dell'istruzione universitaria potendo l'utente impostare anche le informazioni inerenti l'università che frequenta e gli esami che ha sostenuto di modo che se un utente ha difficolta con un esame può ricercare le persone del suo stesso ateneo e corso di studi che lo hanno già sostenuto per chiedere una mano.

I giudizi sulle foto, l'invio e l'accettazione delle richieste di amicizia sono notificate sul cellulare(nel caso di cellulare spento o internet disattivato la notifica arriva appena il cellulare torna reperibile).

Plaze è un applicazione android sviluppata con eclipse ADT un IDE utilissimo che oltre a facilitare la scrittura di codice con suggerimenti e auto completamenti, permette anche una semplice ed intuitiva costruzione di layout grafici, grazie alla quale sono riuscito a studiare la grafica di android semplicemente usandola.

La chat e le notifiche si basano su Google Cloud Messaging permettendo così la ricezione di dati anche se l'applicazione non è in funzione.

Come database è stato usato PostgresSQL con l'estensione postGIS per calcolare la distanza tra coordinate. Plaze è un'applicazione con architettura client - server a tre livelli scritta in linguaggio JAVA con tecnologia ANDROID.

1. Il livello server, è composto da due applicazioni; una per gestire il server vero e proprio, login, registrazione, posizione utente ecc. l'altra per gestire la chat e le notifiche.

2. Il livello client, che sarebbe l'app vera e propria.

3. il terzo livello è la base dati, il data base usato è Postgresql con estensione PostGis per la georeferenziazione.

# NuSeen: an eclipse-based environment for the NuSMV model checker

Paolo Arcaini
CNR – IDPA
paolo.arcaini@idpa.cnr.it

Angelo Gargantini
University of Bergamo
angelo.gargantini@unibg.it

Paolo Vavassori
University of Bergamo
paolo.vavassori@unibg.it

## 1. INTRODUCTION

NuSMV [5] is a symbolic model checker originated from the reengineering, reimplementation and extension of CMU SMV, the original BDD-based model checker developed at CMU by McMillan [4]. The NuSMV project aims at the development of a state-of-the-art symbolic model checker, designed to be applicable in technology transfer projects: it is a well structured, open, flexible and documented platform for model checking, and is robust and close to industrial systems standards [3].

NuSMV has a rich and powerful language that can be used to describe complex systems, which contain the specification of the system behavior as Finite State Machines and its expected requirements (often given by temporal formula).

It can also be used as a model checker, both as a BDD-based symbolic model checker, and as a Bounded model checker. It has around 50 options when it is called in a batch mode. It is widely used as back end for the verification of properties of systems given by means of other formal notations (as for the Abstract State Machines in [1]).

In [2], we have developed a model advisor for NuSMV models. The model advisor performs automatic review of NuSMV models, with the aim of determining if a model is of sufficient quality, where quality is measured as the absence of certain faults. Vulnerabilities and defects a developer can introduce during the modeling activity using NuSMV are expressed as the violation of formal meta-properties. These meta-properties are then mapped to temporal logic formulas, and the NuSMV model-checker itself is used as the engine of our model advisor to notify meta-properties violations, so revealing the absence of some quality attributes of the specification.

## 2. NUSEEN: A NUSMV ECLIPSE-BASED ENVIRONMENT

NuSeen is an eclipse-based environment for NuSMV, with the aim of helping NuSMV users. It mainly focuses in easing the use of the NuSMV tool by means of graphical elements like buttons, menu, text highlighting, and so on. It features:

- A *language* defined by a grammar (concrete syntax) and provided with metamodel (abstract syntax)
- An *editor* that can be used to write NuSMV models and provides an useful feedback like syntax highlighting, autocompletion, and outline.
- A way to *execute* the NuSMV model checker inside eclipse.
- An integrated version of the model advisor which can be executed in eclipse.

### 2.1 Language and editor

NuSeen introduces the definition of the language and of the editor for NuSMV by Xtext [6]. Xtext is a framework for development of programming languages and domain specific languages. It covers all aspects of a complete language infrastructure, from parsers, over linker, compiler or interpreter to fully-blown top-notch Eclipse IDE integration. It comes with good defaults for all these aspects and at the same time every single aspect can be tailored to your needs. We have defined a grammar for the NuSMV notation in the Xtext format. From that, we automatically obtain:

1. a *meta-model* for the language in the form of an EMF (eclipse metamodeling framework) model in the ecore format. It represents the *abstract* syntax of our language in terms of classes and relations. From the meta-model, we automatically obtain the Java APIs that are able to manage (query, create, and modify) NuSMV models and parts of them. This allows the integration of NuSMV with other Java-based tools.

2. an *editor* with *syntax coloring* based on the lexical structure, *content assist* that proposes valid code completions at any place in the document, helping users with the syntactical details of the language, *validation and quick fixes*, linking errors, the outline view, find references, etc.

The editor can be easily extended in order to support extra semantics validation rules, particular rules for indentation and outlining, and other ad hoc editing rules.

### 2.2 NuSMV executor

NuSeen allows the user to run NuSMV from within eclipse. We exploit the *launching framework* in Eclipse. We have performed the following steps:

1. The first step consists in declaring a config type for NuSMV models, by extending the non-ui extension point `launchConfigurationTypes` in the (non-ui) package `org.eclipse.debug.core`.

2. The extension is implemented in a class that actually executes NuSMV with the model chosen by the user.

3. We then prepare the launch configuration dialog (often called LCD) by defining an extension for the extension

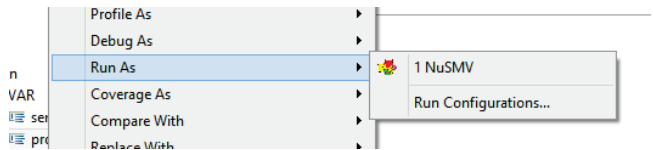

Figure 1: The launch shortcut for NuSMV

point `launchConfigurationTabGroups` in the package `org.eclipse.debug.ui`. It contains the configuration with all the options a user can select when launching NuSMV.

4. We than declare and implement a launch shortcut that allows users to identify a resource in the workbench (either via selection or the active editor) and launch that resource with a single click without bringing up the launch configuration dialog. An example of its use is shown in Fig. 1.

### 2.3 Model Advisor

The model advisor it analyzes a NuSMV specification using NuSMV itself. For a detailed theoretical motivation of the kind of review performed by the model advisor, please see [2]. The NuSMV models are read and queried by the model advisor by using the API generated from the metamodel introduced by the language component. The model adivsor runs NuSMV using the APIs defined in the executor. To run the model advisor from the eclipse UI, we exploit again the launching framework of eclipse. We

## 3. ARCHITECTURE

NuSeen is divided in three main components, as shown in Fig. 2, each divided in one or more plugin eclipse projects:

1. The component containing the definition of the language and the editor (dsl by xtext). It is divided into two plugins: one contains the definition of the grammar and the other the classes for the editor.

2. The component containing the runner, which simply introduces the launching framework for the execution of NuSMV.

3. The model advisor which contains the model advisor itself, its UI part in order to ingrate it within eclipse and NuSeen, and a third project which allows the user to call the model advisor form outside eclipse. This latter plugin wraps everything needed by the model advisor in an executable jar file.

## 4. FUTURE WORK

We plan to maintain and extend NuSeen in the following directions. About the language, we would like to test it and to add more precise semantic rules. Up to now, the editor can still consider correct files that are actually invalid according to the real NuSMV semantics and that can be checked only by the NuSMV parser. We plan to integrate the execution of NuSMV inside eclipse, by releasing the user from the burden of installing and configuring NuSMV. Up to now, we assume that the user has already installed NuSMV (which must be in the path). We plan to use technologies like JNA to ship NuSMV together with NuSeen.

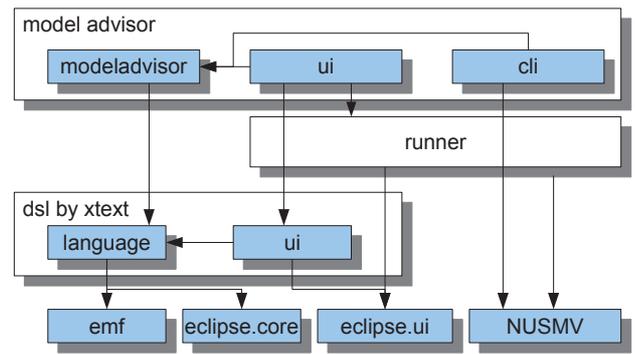

Figure 2: Architecture

The feedback given by the NuSMV parser and by the model advisor is simply shown in the console. We plan to read possible errors and warnings found by NuSMV and by the advisor and show them directly in the editor by using appropriate markers.

## 5. ACKNOWLEDGEMENTS

We would like to thank Siamak Haschemi for the initial version of the XTEXT for NuSMV grammar.

## 6. REFERENCES

[1] P. Arcaini, A. Gargantini, and E. Riccobene. AsmetaSMV: a way to link high-level ASM models to low-level NuSMV specifications. In M. Frappier, U. Glässer, S. Khurshid, R. Laleau, and S. Reeves, editors, *ABZ 2010*, volume 5977 of *Lecture Notes in Computer Science*, pages 61–74. Springer, 2010.
[2] P. Arcaini, A. Gargantini, and E. Riccobene. A model advisor for NuSMV specifications. *Innovations in Systems and Software Engineering*, 7:97–107, 2011.
[3] A. Cimatti, E. Clarke, F. Giunchiglia, and M. Roveri. NUSMV: a new symbolic model checker. *International Journal on Software Tools for Technology Transfer (STTT)*, 2(4):410–425, Mar. 2000.
[4] K. L. McMillan. The SMV system, symbolic model checking - an approach. Technical Report CMU-CS-92-131, Carnegie Mellon University, 1992.
[5] The NuSMV website. `http://nusmv.fbk.eu/`.
[6] Xtext. http://www.eclipse.org/xtext/.